\documentclass[runningheads,a4paper]{llncs}

\usepackage{todonotes}
\usepackage[utf8]{inputenc}
\usepackage{url}
\usepackage{alltt}
\usepackage{amssymb,amsmath}
\usepackage[english]{babel}
\usepackage{mathpartir}
\usepackage{isabelle,isabellesym}
\isabellestyleit
\usepackage{caption}
\usepackage{subcaption}
\usepackage{proof}
\usepackage{semantic}

\usepackage{tikz}
\usetikzlibrary{trees}
\usetikzlibrary{arrows}
\usetikzlibrary{decorations.pathmorphing}
\usetikzlibrary{shapes.multipart}
\usetikzlibrary{shapes.geometric}
\usetikzlibrary{calc}
\usetikzlibrary{positioning} 
\usetikzlibrary{fit}
\usetikzlibrary{backgrounds}

\usepackage{tikz-cd}

\usepackage[colorlinks,hyperindex,bookmarks,linkcolor=blue,citecolor=blue,urlcolor=blue]{hyperref}

\newcommand{\secref}[1]{Section~\ref{#1}}

\newcommand{\figref}[1]{Figure~\ref{#1}}

\newcommand{\lemmaref}[1]{Lemma~\ref{#1}}

\newcommand{\figline}{\rule{\textwidth}{1.0pt}}

\newcommand{\remms}[2][]{}
\newcommand{\remnb}[2][]{}
\newcommand{\remwr}[2][]{}

\definecolor{darkred}{rgb}{0.7,0,0}

\newcommand{\beforeexec}{\downarrow\hspace{-0.6ex}}
\newcommand{\afterexec}{\hspace{-0.8ex}\uparrow}

\begin{document}

\mainmatter

\title{Abstracting an operational semantics to finite automata}
\author{Nadezhda Baklanova, 
  Wilmer Ricciotti, 
  Jan-Georg Smaus,
  Martin Strecker
}
\institute{
  IRIT (Institut de Recherche en Informatique de Toulouse)\\
  Universit{\'e} de Toulouse, France\\
  \texttt{\emph{firstname}.\emph{lastname}@irit.fr}
  \thanks{N.~Baklanova and M.~Strecker were partially supported by the project \emph{Verisync} (ANR-10-BLAN-0310). }
  \fnmsep
  \thanks{W.~Ricciotti and J.-G. Smaus are supported by the project \textsc{Ajitprop} of the Fondation Airbus.} 
}

\maketitle
\thispagestyle{empty}
\pagestyle{empty}

\begin{abstract}
There is an apparent similarity between the descriptions of small-step
operational semantics of imperative programs and the semantics of finite
automata, so defining an abstraction mapping from semantics to automata and
proving a simulation property seems to be easy. This paper aims at identifying
the reasons why simple proofs break, among them artifacts in the semantics
that lead to stuttering steps in the simulation. We then present a semantics
based on the zipper data structure, with a direct interpretation of evaluation
as navigation in the syntax tree. The abstraction function is then defined by
equivalence class construction. 


\end{abstract}

\section{Introduction}\label{sec:introduction}

Among the formalisms employed to describe the semantics of transition systems,
two particularly popular choices are abstract machines and structural
operational semantics (SOS). Abstract machines are widely used for modeling and
verifying dynamic systems, e.g. finite automata, B\"{u}chi automata or timed
automata \cite{Khoussainov:2001:ATA:558914,Baier2008,DBLP:conf/lics/AlurCD90}.
An abstract machine can be represented as a directed graph with transition
semantics between nodes.  The transition semantics is defined by moving a pointer
to a current node.  Automata are a popular tool for modeling dynamic systems due
to the simplicity of the verification of automata systems, which can be carried
out in a fully automated way, something that is not generally possible for
Turing-complete systems.

This kind of semantics is often extended by adding a background state composed
of a set of variables with their values: this is the case of timed automata,
which use background clock variables \cite{Alur94atheory}. The \textsc{Uppaal}
model checker for timed automata extends the notion of background state even
further by adding integer and Boolean variables to the state
\cite{springerlink:10.1007/978-3-540-27755-2} which, however, do not increase
the computational power of such timed automata but make them more convenient to
use.


Another formalism for modeling transition systems is structural semantics
(``small-step'', contrary to ``big-step'' semantics which is much easier to
handle but which is inappropriate for a concurrent setting), which uses a set
of reduction rules for simplifying a program expression. It has been described
in detail in \cite{Winskel:1993:FSP:151145} and used, for example, for the
Jinja project developing a formal model of the Java language
\cite{Klein:2006:MMJ:1146809.1146811}. An appropriate semantic rule for
reduction is selected based on the expression pattern and on values of some
variables in a state. As a result of reduction the expression and the state
are updated.

$$\inference*{s'=s(v\longmapsto eval\; expr\; s)}{(Assign\; v\;
expr,s)\rightarrow(Unit,s')}[\textsc{[Assignment]}]$$

This kind of rules is intuitive; however, the proofs involving them require
induction over the expression structure. A different approach to writing a
structural semantics was described in
\cite{Appel07separationlogic,leroy_compcert-backend_jar_2009} for the CMinor
language. It uses a notion of continuation which represents an expression as a
control stack and deals with separate parts of the control stack consecutively.

$$(Seq\;e1\;e2\cdot\kappa,s)\rightarrow(e1\cdot e2\cdot \kappa,s)\quad \quad
\quad (Empty\cdot\kappa,s)\rightarrow(\kappa,s)$$

Here the ``$\cdot$'' operator designates concatenation of control stacks. The
semantics of continuations does not need induction over the expression,
something which makes proof easier; however it requires more auxiliary steps for
maintaining the control stack which do not have direct correspondance in the
modeled language.

For modeling non-local transfer of control, Krebbers and Wiedijk
\cite{krebbers13:_separ_logic_non_contr_flow} present a semantics using
(non-recursive) ``statement contexts''. These are combined with the
above-mentioned continuation stacks. The resulting semantics is situated
mid-way between \cite{Appel07separationlogic} and the semantics proposed
below.

The present paper describes an approach to translation from structural
operational semantics to finite automata extended with background state. All
the considered automata are an extension of B\"{u}chi automata with background
state, i.e. they have a finite number of nodes and edges but can produce an
infinite trace. The reason of our interest in abstracting from structural
semantics to B\"uchi automata is our work in progress
\cite{baklanova12:_abstr_verif_proper_real_time_java}. We are working on a
static analysis algorithm for finding possible resource sharing conflicts in
multithreaded Java programs. For this purpose we annotate Java programs with
timing information and then translate them to a network of timed automata
which is later model checked. The whole translation is formally verified. One
of the steps of the translation procedure includes switching from structural
operational semantics of a Java-like language to automata semantics. During
this step we discovered some problems which we will describe in the next
section. The solutions we propose extend well beyond the problem of
abstracting a structured language to an automaton. It can also be used for
compiler verification, which usually is cluttered up with arithmetic adress
calculation that can be avoided in our approach.

The contents of the paper has been entirely formalized in the
Isabelle proof assistant \cite{Isabelle_Tutorial}. We have not insisted on any
Isabelle-specific features, therefore this formalization can be rewritten using
other proof assistants. The full Isabelle formal development can
be found on the web \cite{zipper_formalization}.


 \section{Problem Statement}\label{sec:problem_statement}

We have identified the following as the main problems when trying to prove the
correctness of the translation between a programming language semantics and
its abstraction to automata:

\begin{enumerate}
\item Preservation of execution context: an abstract
machine always sees all the available nodes while a reduced expression loses
the information about previous reductions. \label{fst_problem}
\item Semantic artifacts: some reduction rules are necessary for the functionality of the semantics,
  but may be missing in the modeled language. Additionally, the rules can
  produce expressions which do not occur in the original language. \label{snd_problem}
\end{enumerate}

These problems occur independently of variations in the presentation of
semantic rules \cite{Winskel:1993:FSP:151145}  adopted in the literature, such as
\cite{Klein:2006:MMJ:1146809.1146811} (recursive evaluation of sub-statements)
or \cite{Appel07separationlogic,leroy_compcert-backend_jar_2009} (continuation-style).

\begin{isabellebody}%
\def\isabellecontext{Zipper}%
\isadelimtheory
\endisadelimtheory
\isatagtheory
\endisatagtheory
{\isafoldtheory}%
\isadelimtheory
\endisadelimtheory
\isamarkuptrue%
\begin{isamarkuptext}%
We will describe these two problems in detail, and later our approach to their
solution, in the context of a minimalistic programming language which only
manipulates Boolean values (a \isa{Null} value is also added to account for
errors):%
\end{isamarkuptext}%
\isamarkuptrue%
\isacommand{datatype}\isamarkupfalse%
\ val\ {\isacharequal}\ Bool\ bool\ {\isacharbar}\ Null%
\begin{isamarkuptext}%
The language can be extended in a rather straightforward way to more complex expressions.
In this language, expressions are either values or variables:%
\end{isamarkuptext}%
\isamarkuptrue%
\isacommand{datatype}\isamarkupfalse%
\ expr\ {\isacharequal}\ Val\ val\ {\isacharbar}\ Var\ vname%
\begin{isamarkuptext}%
The statements are those of a small imperative language (similarly to 
  \cite{nipkow_klein_concrete_semantics_2013}):%
\end{isamarkuptext}%
\isamarkuptrue%
\isacommand{datatype}\isamarkupfalse%
\ stmt\ {\isacharequal}\ \isanewline
\ \ \ \ Empty\ \ \ \ \ \ \ \ \ \ \ \ \ \ \ \ \ %
\isamarkupcmt{no-op%
}
\isanewline
\ \ {\isacharbar}\ Assign\ vname\ val\ \ \ \ \ \ %
\isamarkupcmt{assignment: $var$ := $val$%
}
\isanewline
\ \ {\isacharbar}\ Seq\ stmt\ stmt\ \ \ \ \ \ \ \ \ \ \ %
\isamarkupcmt{sequence: $c_1; c_2$%
}
\isanewline
\ \ {\isacharbar}\ Cond\ expr\ stmt\ stmt\ \ \ \ %
\isamarkupcmt{conditional: if $e$ then $c_1$ else $c_2$%
}
\isanewline
\ \ {\isacharbar}\ While\ expr\ stmt\ \ \ \ \ \ \ \ \ \ %
\isamarkupcmt{loop:  while $e$ do $c$%
}
\end{isabellebody}
\medskip

\subsection{Preservation of execution context}
Problem~\ref{fst_problem} concerns the loss of an execution context
through expression reductions which is a design feature of structural
semantics.  Let us consider a simple example. 

\begin{figure}[hb]
$$\inference*{s'=s(v\longmapsto eval\; expr\; s)}{(Assign\; v\;
  expr,s)\rightarrow(Empty,s')}[\textsc{[Assign]}] \quad$$
$$\inference*{eval\; bexp\; s=True}{(Cond\;bexp\;e1\;
  e2,s)\rightarrow(e1,s)}[\textsc{[CondT]}] \quad
\inference*{eval\; bexp\; s=False}{(Cond\;bexp\;e1\;
  e2,s)\rightarrow(e2,s)}[\textsc{[CondF]}]$$
\caption{Semantic rules for the minimal imperative language.}\label{fig:sosrules}
\end{figure}

Assume we have a structural semantics for our minimal imperative language (some
rules of a traditional presentation are shown in \figref{fig:sosrules}): we want to
translate a program written in this language into an abstract machine. Assume
that the states of variable values have the same representation in the two systems:
this means we only need to translate the program expression into a directed
graph with different nodes corresponding to different expressions obtained by
reductions of the initial program expression.

On the abstract machine level the $Assign$ statements would be represented as two-state automata, and
the $Cond$ as a node with two outgoing edges directed to the automata for the
bodies of its branches.

Consider a small program in this language $Cond\;bexp\;(Assign\;a\;5)\;Empty$ and its
execution flow.

{\small
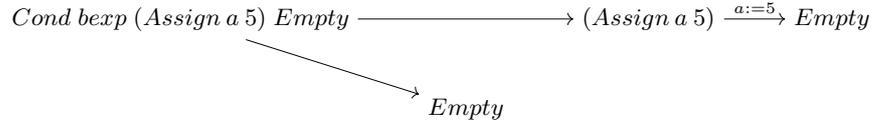
\begin{tikzcd}
Cond \;bexp\; (Assign\:a\:5)\; Empty \arrow{rr}{} \arrow{rd}{} & &(Assign\:a\:5) \arrow{r}{a:=5} &Empty \\
&  Empty  & & 
\end{tikzcd}
}

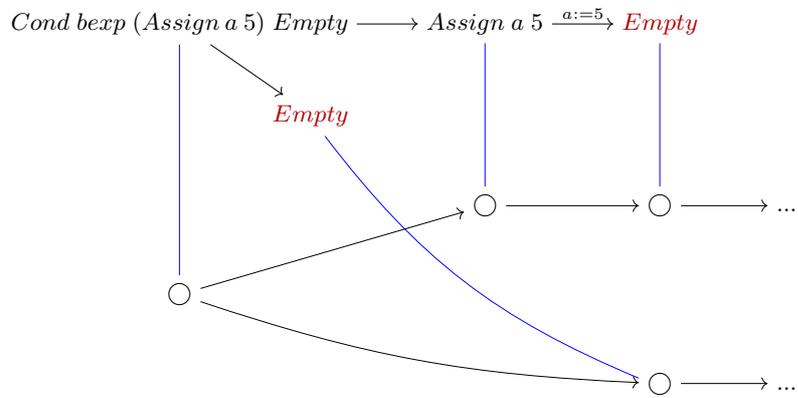
\begin{figure}
{\footnotesize
\begin{tikzcd}
Cond \;bexp\; (Assign\: a\: 5)\; Empty \arrow[dash,blue]{ddd}{} \arrow{r}{}
\arrow[start anchor={[xshift=3ex]},end anchor={[xshift=10ex]}]{d}{}
&Assign\; a\; 5 \arrow[dash,blue]{dd}{} \arrow{r}{a:=5} &{\color{darkred} Empty}
\arrow[dash,blue]{dd}{} & \\
{\color{white}Cond \;bexp\; (Assign\: a\: 5) \;} {\color{darkred} Empty}
\arrow[dash,start anchor ={[xshift=12ex]},bend right=15,blue]{rrddd} & & & \\
& \bigcirc \arrow{r}{} &\bigcirc \arrow{r}{} &... \\
 \bigcirc \arrow{ru}{} \arrow[bend right=8]{rrd}{} & & & \\
& &\bigcirc \arrow{r}{} & ...
\end{tikzcd}
}
\caption{The execution flow and the corresponding abstract machine for the
  program $Cond\;bexp\;(Assign\:a\:5)\;Empty$.}
\label{collision_example}
\end{figure}

The execution can select any of the two branches depending on the $bexp$
value. 
There are two different $Empty$ expressions appearing as results of two
different reductions. The corresponding abstract machine would be a natural
graph representation for a condition statement with two branches (\figref{collision_example}).

During the simple generation of an abstract machine from a program expression the
two $Empty$ statements cannot be distinguished although they should be mapped
into two different nodes in the graph. We need to add more information about
the context into the translation, and it can be done by different ways. 

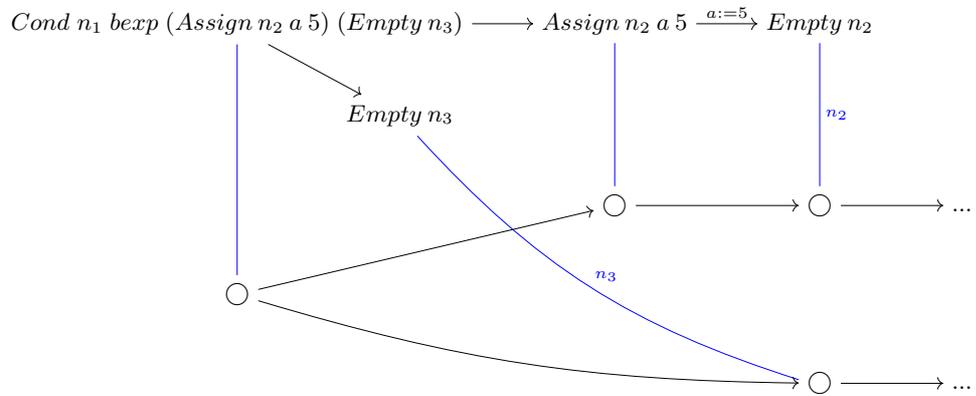
\begin{figure}
{\footnotesize
\begin{tikzcd}
Cond\; n_1 \;bexp\; (Assign\:n_2\: a\: 5)\; (Empty\:n_3) \arrow[dash,blue]{ddd}{} \arrow{r}{}
\arrow[start anchor={[xshift=3ex]},end anchor={[xshift=12ex]}]{d}{}
&Assign\:n_2\: a\: 5 \arrow[dash,blue]{dd}{} \arrow{r}{a:=5} &Empty\:n_2
\arrow[dash,blue]{dd}{n_2} & \\
{\color{white}Cond\; n_1 \;bexp\; (Assign\:n_2\: a\: 5)\;} Empty\: n_3
\arrow[dash,start anchor ={[xshift=15ex]},bend right=15,blue]{rrddd}{n_3} & & & \\
& \bigcirc \arrow{r}{} &\bigcirc \arrow{r}{} &... \\
 \bigcirc \arrow{ru}{} \arrow[bend right=8]{rrd}{} & & & \\
& &\bigcirc \arrow{r}{} & ...
\end{tikzcd}
}
\caption{The execution flow and the corresponding abstract machine for the
  program with subexpression identifiers
  $Cond\;n_1\;bexp\;(Assign\:n_2\:a\:5)\;(Empty\:n_3)$.}
\label{collision_example_ids}
\end{figure}

A straightforward solution would be to add some information in order to
distinguish between the two $Empty$ expressions. If we add unique identifiers
to each subexpression of the program, they will allow to know exactly which subexpression  we
are translating (Figure~\ref{collision_example_ids}).
The advantage of this approach is its simplicity, however, it requires
additional  functions and proofs for identifier management.

Another solution for the problem proposed in this paper involves usage of
a special data
structure to keep the context of the translation.
There are known examples of translations from subexpression-based semantics 
\cite{Klein:2006:MMJ:1146809.1146811} and continuation-based semantics
\cite{leroy_compcert-backend_jar_2009} to abstract
machines. However, all these translations do not address the problem of
 context preservation during the translation.

\subsection{Semantic artifacts}
The second problem appears because of the double functionality of the
$Empty$ expression: it is used to define an empty operator which does nothing
as well as the final expression for reductions which cannot be further
reduced. The typical semantic rules for a sequence of expressions look as 
shown on \figref{seqsem}.

\begin{figure}
$$\inference*{(e1,s)\rightarrow(e1',s')}{(Seq\; e1\;
  e2,s)\rightarrow(Seq\;e1'\;e2,s')}[\textsc{[Seq1]}] \quad
\inference*{}{(Seq\;Empty\;e2,s)\rightarrow(e2,s)}[\textsc{[Seq2]}]$$
\caption{Semantic rules for the sequence of two expressions.}\label{seqsem}
\end{figure}

Here the $Empty$ expression means that the first expression in the sequence
has been reduced up to the end, and we can start reducing the second
expression. However, any imperative language translated to an assembly
language would not have an additional operator between the two pieces of code
corresponding to the first and the second expressions. The rule
\textsc{Seq2} must be marked as a silent transition when translated to
an automaton, or the semantic rules have to be changed.


%
\begin{isabellebody}%
\def\isabellecontext{Zipper}%
\isadelimtheory
\endisadelimtheory
\isatagtheory
\endisatagtheory
{\isafoldtheory}%
\isadelimtheory
\endisadelimtheory
\isamarkupsection{Zipper-based semantics of imperative programs\label{sec:source}%
}
\isamarkuptrue%
\isamarkupsubsection{The zipper data structure\label{sec:source_syntax}%
}
\isamarkuptrue%
\begin{isamarkuptext}%
Our plan is to propose an alternative technique to formalize operational 
semantics that will make it easier to preserve the execution context during
the translation to an automata-based formalism. Our technique is built around a zipper
data structure, whose purpose is to identify a location in a tree (in our
case: a \isa{stmt}) by the subtree below the location and the rest of the
tree (in our case: of type \isa{stmt{\isacharunderscore}path}).  In order to allow for an easy
navigation, the rest of the tree is turned inside-out so that it is possible
to reach the root of the tree by following the backwards pointers.
The following definition is a straightforward adaptation of the zipper for
binary trees discussed in \cite{Huet_zipper:1997} to the \isa{stmt} data
type:%
\end{isamarkuptext}%
\isamarkuptrue%
\isacommand{datatype}\isamarkupfalse%
\ stmt{\isacharunderscore}path\ {\isacharequal}\ \isanewline
\ \ PTop\isanewline
{\isacharbar}\ PSeqLeft\ stmt{\isacharunderscore}path\ stmt\ \ \ \ \ \ \ \ \ \ {\isacharbar}\ PSeqRight\ stmt\ stmt{\isacharunderscore}path\isanewline
{\isacharbar}\ PCondLeft\ expr\ stmt{\isacharunderscore}path\ stmt\ \ {\isacharbar}\ PCondRight\ expr\ stmt\ stmt{\isacharunderscore}path\isanewline
{\isacharbar}\ PWhile\ expr\ stmt{\isacharunderscore}path%
\begin{isamarkuptext}%
Here, \isa{PTop} represents the root of the original tree, and for
each constructor of \isa{stmt} and each of its sub-\isa{stmt}s, there is a
``hole'' of type \isa{stmt{\isacharunderscore}path} where a subtree can be fitted in. A
location in a tree is then a combination of a \isa{stmt} and a \isa{stmt{\isacharunderscore}path}:%
\end{isamarkuptext}%
\isamarkuptrue%
\isacommand{datatype}\isamarkupfalse%
\ stmt{\isacharunderscore}location\ {\isacharequal}\ Loc\ stmt\ stmt{\isacharunderscore}path%
\begin{isamarkuptext}%
Given a location in a tree, the function \isa{reconstruct} reconstructs the
 original tree \isa{reconstruct\ {\isacharcolon}{\isacharcolon}\ stmt\ {\isasymRightarrow}\ stmt{\isacharunderscore}path\ {\isasymRightarrow}\ stmt}, and \isa{reconstruct{\isacharunderscore}loc\ {\isacharparenleft}Loc\ c\ sp{\isacharparenright}\ {\isacharequal}\ reconstruct\ c\ sp} does the same for a location.%
\end{isamarkuptext}%
\isamarkuptrue%
\isacommand{fun}\isamarkupfalse%
\ reconstruct\ {\isacharcolon}{\isacharcolon}\ {\isachardoublequoteopen}stmt\ {\isasymRightarrow}\ stmt{\isacharunderscore}path\ {\isasymRightarrow}\ stmt{\isachardoublequoteclose}\ \isakeyword{where}\isanewline
\ \ {\isachardoublequoteopen}reconstruct\ c\ PTop\ {\isacharequal}\ c{\isachardoublequoteclose}\isanewline
{\isacharbar}\ {\isachardoublequoteopen}reconstruct\ c\ {\isacharparenleft}PSeqLeft\ sp\ c{\isadigit{2}}{\isacharparenright}\ {\isacharequal}\ reconstruct\ {\isacharparenleft}Seq\ c\ c{\isadigit{2}}{\isacharparenright}\ sp{\isachardoublequoteclose}\isanewline
{\isacharbar}\ {\isachardoublequoteopen}reconstruct\ c\ {\isacharparenleft}PSeqRight\ c{\isadigit{1}}\ sp{\isacharparenright}\ {\isacharequal}\ reconstruct\ {\isacharparenleft}Seq\ c{\isadigit{1}}\ c{\isacharparenright}\ sp{\isachardoublequoteclose}\isanewline
{\isacharbar}\ {\isachardoublequoteopen}reconstruct\ c\ {\isacharparenleft}PCondLeft\ e\ sp\ c{\isadigit{2}}{\isacharparenright}\ {\isacharequal}\ reconstruct\ {\isacharparenleft}Cond\ e\ c\ c{\isadigit{2}}{\isacharparenright}\ sp{\isachardoublequoteclose}\isanewline
{\isacharbar}\ {\isachardoublequoteopen}reconstruct\ c\ {\isacharparenleft}PCondRight\ e\ c{\isadigit{1}}\ sp{\isacharparenright}\ {\isacharequal}\ reconstruct\ {\isacharparenleft}Cond\ e\ c{\isadigit{1}}\ c{\isacharparenright}\ sp{\isachardoublequoteclose}\isanewline
{\isacharbar}\ {\isachardoublequoteopen}reconstruct\ c\ {\isacharparenleft}PWhile\ e\ sp{\isacharparenright}\ {\isacharequal}\ reconstruct\ {\isacharparenleft}While\ e\ c{\isacharparenright}\ sp{\isachardoublequoteclose}\isanewline
\isanewline
\isanewline
\isacommand{fun}\isamarkupfalse%
\ reconstruct{\isacharunderscore}loc\ {\isacharcolon}{\isacharcolon}\ {\isachardoublequoteopen}stmt{\isacharunderscore}location\ {\isasymRightarrow}\ stmt{\isachardoublequoteclose}\ \isakeyword{where}\isanewline
\ \ {\isachardoublequoteopen}reconstruct{\isacharunderscore}loc\ {\isacharparenleft}Loc\ c\ sp{\isacharparenright}\ {\isacharequal}\ reconstruct\ c\ sp{\isachardoublequoteclose}\isanewline
\isamarkupsubsection{Semantics\label{sec:source_sem}%
}
\isamarkuptrue%
\begin{isamarkuptext}%
Our semantics is a small-step operational semantics describing the
effect of the execution a program on a certain program state. For each
variable, the state yields \isa{Some} value associated with the variable, or
\isa{None} if the variable is unassigned. More formally, the state is a
mapping \isa{vname\ {\isasymRightarrow}\ val\ option}.  Defining the evaluation of an
expression in a state is then standard. 

Before commenting the rules of our semantics, let us discuss which kind
of structure we are manipulating. The semantics essentially consists in moving
around a pointer within the syntax tree. As explained in \secref{sec:source_syntax}, a
position in the syntax tree is given by a \isa{stmt{\isacharunderscore}location}. However,
during the traversal of the syntax tree, we visit each position at least twice
(and possibly several times, for example in a loop): before executing the
corresponding statement, and after finishing the execution. We therefore add a
Boolean flag, where \isa{True} is a marker for ``before'' and \isa{False}
for ``after'' execution. 

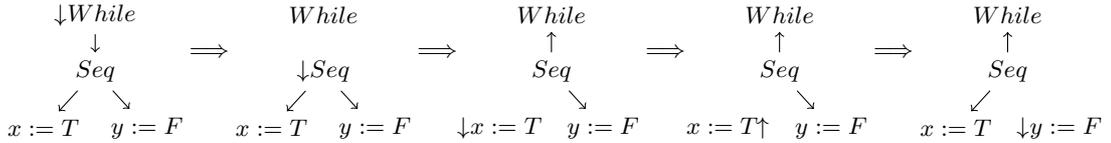
\begin{figure}
\hspace{-10ex}
\begin{tikzpicture}[scale=0.5,level/.style={sibling distance = 20ex}]
\node at (0, 0) {$\beforeexec While$}
child {node {$Seq$}
  child {
    node {$x := T$}
    edge from parent [->]
  }
  child {
    node {$y := F$}
    edge from parent [->]
  }
  edge from parent [->]
};

\node at (3, -1) {$\Longrightarrow$};

\node at (6, 0) {$While$}
child {node {$\beforeexec Seq$}
  child {
    node {$x := T$}
    edge from parent [->]
  }
  child {
    node {$y := F$}
    edge from parent [->]
  }
  edge from parent [draw=none]  
};

\node at (9, -1) {$\Longrightarrow$};

\node at (12, 0) {$While$}
child {node {$Seq$}
  child {
    node {$\beforeexec x := T$}
    edge from parent [draw=none]
  }
  child {
    node {$y := F$}
    edge from parent [->]
  }
  edge from parent [<-]  
};

\node at (15, -1) {$\Longrightarrow$};

\node at (18, 0) {$While$}
child {node {$Seq$}
  child {
    node {$x := T\afterexec$}
    edge from parent [draw=none]
  }
  child {
    node {$y := F$}
    edge from parent [->]
  }
  edge from parent [<-]  
};

\node at (21, -1) {$\Longrightarrow$};

\node at (24, 0) {$While$}
child {node {$Seq$}
  child {
    node {$x := T$}
    edge from parent [->]
  }
  child {
    node {$\beforeexec y := F$}
    edge from parent [draw=none]
  }
  edge from parent [<-]  
};

\end{tikzpicture}

\caption{Example of execution of small-step semantics}
\label{fig:example_execution_semantics}
\end{figure}

As an example, consider the execution sequence depicted in
\figref{fig:example_execution_semantics} (with assignments written in a more readable concrete syntax), consisting of the initial steps of
the execution of the program \isa{While\ {\isacharparenleft}e{\isacharcomma}\ Seq{\isacharparenleft}x\ {\isacharcolon}{\isacharequal}\ T{\isacharcomma}\ y\ {\isacharcolon}{\isacharequal}\ F{\isacharparenright}{\isacharparenright}}. The before
(resp.{} after) marker is indicated by a downward arrow before (resp.{} an upward
arrow behind) the current statement. 
The condition of the loop is omitted because it is irrelevant here. 
The middle configuration would be coded as
\isa{{\isacharparenleft}{\isacharparenleft}Loc\ {\isacharparenleft}x\ {\isacharcolon}{\isacharequal}\ T{\isacharparenright}\ {\isacharparenleft}PSeqLeft\ {\isacharparenleft}PWhile\ e\ PTop{\isacharparenright}\ {\isacharparenleft}y\ {\isacharcolon}{\isacharequal}\ F{\isacharparenright}{\isacharparenright}{\isacharparenright}{\isacharcomma}\ True{\isacharparenright}}.

Altogether, we obtain a syntactic configuration (\isa{synt{\isacharunderscore}config}) which
combines the location and the Boolean flag. The semantic configuration (\isa{sem{\isacharunderscore}config}) manipulated by the semantics adjoins the \isa{state}, as
defined previously.%
\end{isamarkuptext}%
\isamarkuptrue%
\isacommand{type{\isacharunderscore}synonym}\isamarkupfalse%
\ synt{\isacharunderscore}config\ {\isacharequal}\ {\isachardoublequoteopen}stmt{\isacharunderscore}location\ {\isasymtimes}\ bool{\isachardoublequoteclose}\isanewline
\isacommand{type{\isacharunderscore}synonym}\isamarkupfalse%
\ sem{\isacharunderscore}config\ {\isacharequal}\ {\isachardoublequoteopen}synt{\isacharunderscore}config\ {\isasymtimes}\ state{\isachardoublequoteclose}%
\begin{isamarkuptext}%
The rules of the small-step semantics of \figref{fig:smallStep} fall
into two categories: before execution of a statement \isa{s} (of the form \isa{{\isacharparenleft}{\isacharparenleft}l{\isacharcomma}\ True{\isacharparenright}{\isacharcomma}\ s{\isacharparenright}}) and after execution (of the form \isa{{\isacharparenleft}{\isacharparenleft}l{\isacharcomma}\ False{\isacharparenright}{\isacharcomma}\ s{\isacharparenright}});
there is only one rule of this latter kind: {\sc SFalse}.%
\end{isamarkuptext}%
\isamarkuptrue%
\begin{figure}[h!]
\isastyleminor\isamarkuptrue
\figline{}
\isacommand{fun}\isamarkupfalse%
\ next{\isacharunderscore}loc\ {\isacharcolon}{\isacharcolon}\ {\isachardoublequoteopen}stmt\ {\isasymRightarrow}\ stmt{\isacharunderscore}path\ {\isasymRightarrow}\ {\isacharparenleft}stmt{\isacharunderscore}location\ {\isasymtimes}\ bool{\isacharparenright}{\isachardoublequoteclose}\ \isakeyword{where}\isanewline
\ \ {\isachardoublequoteopen}next{\isacharunderscore}loc\ c\ PTop\ {\isacharequal}\ {\isacharparenleft}Loc\ c\ PTop{\isacharcomma}\ False{\isacharparenright}{\isachardoublequoteclose}\isanewline
{\isacharbar}\ {\isachardoublequoteopen}next{\isacharunderscore}loc\ c\ {\isacharparenleft}PSeqLeft\ sp\ c\isactrlsub {\isadigit{2}}{\isacharparenright}\ {\isacharequal}\ {\isacharparenleft}Loc\ c\isactrlsub {\isadigit{2}}\ {\isacharparenleft}PSeqRight\ c\ sp{\isacharparenright}{\isacharcomma}\ True{\isacharparenright}{\isachardoublequoteclose}\isanewline
{\isacharbar}\ {\isachardoublequoteopen}next{\isacharunderscore}loc\ c\ {\isacharparenleft}PSeqRight\ c\isactrlsub {\isadigit{1}}\ sp{\isacharparenright}\ {\isacharequal}\ {\isacharparenleft}Loc\ {\isacharparenleft}Seq\ \ c\isactrlsub {\isadigit{1}}\ c{\isacharparenright}\ sp{\isacharcomma}\ False{\isacharparenright}{\isachardoublequoteclose}\isanewline
{\isacharbar}\ {\isachardoublequoteopen}next{\isacharunderscore}loc\ c\ {\isacharparenleft}PCondLeft\ e\ sp\ c\isactrlsub {\isadigit{2}}{\isacharparenright}\ {\isacharequal}\ {\isacharparenleft}Loc\ {\isacharparenleft}Cond\ e\ c\ c\isactrlsub {\isadigit{2}}{\isacharparenright}\ sp{\isacharcomma}\ False{\isacharparenright}{\isachardoublequoteclose}\isanewline
{\isacharbar}\ {\isachardoublequoteopen}next{\isacharunderscore}loc\ c\ {\isacharparenleft}PCondRight\ e\ c\isactrlsub {\isadigit{1}}\ sp{\isacharparenright}\ {\isacharequal}\ {\isacharparenleft}Loc\ {\isacharparenleft}Cond\ e\ c\isactrlsub {\isadigit{1}}\ c{\isacharparenright}\ sp{\isacharcomma}\ False{\isacharparenright}{\isachardoublequoteclose}\isanewline
{\isacharbar}\ {\isachardoublequoteopen}next{\isacharunderscore}loc\ c\ {\isacharparenleft}PWhile\ e\ sp{\isacharparenright}\ {\isacharequal}\ {\isacharparenleft}Loc\ {\isacharparenleft}While\ e\ c{\isacharparenright}\ sp{\isacharcomma}\ True{\isacharparenright}{\isachardoublequoteclose}%
\newline
\figline{}
\caption{Finding the next location}\label{fig:next_loc}
\end{figure}
\isadelimproof
\endisadelimproof
\isatagproof
\endisatagproof
{\isafoldproof}%
\isadelimproof
\endisadelimproof
\begin{figure}[h!]
\isastyleminor\isamarkuptrue
\figline{}
\begin{isamarkuptext}%
\begin{center}
\begin{small}
\isa{\mbox{}\inferrule{\mbox{}}{\mbox{{\isacharparenleft}{\isacharparenleft}Loc\ Empty\ sp{\isacharcomma}\ True{\isacharparenright}{\isacharcomma}\ s{\isacharparenright}\ {\isasymrightarrow}\ {\isacharparenleft}{\isacharparenleft}Loc\ Empty\ sp{\isacharcomma}\ False{\isacharparenright}{\isacharcomma}\ s{\isacharparenright}}}}{\hspace{1ex} [\sc SEmpty]}\\[2ex]
\isa{\mbox{}\inferrule{\mbox{}}{\mbox{{\isacharparenleft}{\isacharparenleft}Loc\ {\isacharparenleft}Assign\ vr\ vl{\isacharparenright}\ sp{\isacharcomma}\ True{\isacharparenright}{\isacharcomma}\ s{\isacharparenright}\ {\isasymrightarrow}\ {\isacharparenleft}{\isacharparenleft}Loc\ {\isacharparenleft}Assign\ vr\ vl{\isacharparenright}\ sp{\isacharcomma}\ False{\isacharparenright}{\isacharcomma}\ s{\isacharparenleft}vr\ {\isasymmapsto}\ vl{\isacharparenright}{\isacharparenright}}}}{[\sc SAssign]}\\[2ex]
\isa{\mbox{}\inferrule{\mbox{}}{\mbox{{\isacharparenleft}{\isacharparenleft}Loc\ {\isacharparenleft}Seq\ c\isactrlsub {\isadigit{1}}\ c\isactrlsub {\isadigit{2}}{\isacharparenright}\ sp{\isacharcomma}\ True{\isacharparenright}{\isacharcomma}\ s{\isacharparenright}\ {\isasymrightarrow}\ {\isacharparenleft}{\isacharparenleft}Loc\ c\isactrlsub {\isadigit{1}}\ {\isacharparenleft}PSeqLeft\ sp\ c\isactrlsub {\isadigit{2}}{\isacharparenright}{\isacharcomma}\ True{\isacharparenright}{\isacharcomma}\ s{\isacharparenright}}}}{\hspace{1ex} [\sc SSeq]}\\[2ex]
\isa{\mbox{}\inferrule{\mbox{eval\ e\ s\ {\isacharequal}\ Bool\ True}}{\mbox{{\isacharparenleft}{\isacharparenleft}Loc\ {\isacharparenleft}Cond\ e\ c\isactrlsub {\isadigit{1}}\ c\isactrlsub {\isadigit{2}}{\isacharparenright}\ sp{\isacharcomma}\ True{\isacharparenright}{\isacharcomma}\ s{\isacharparenright}\ {\isasymrightarrow}\ {\isacharparenleft}{\isacharparenleft}Loc\ c\isactrlsub {\isadigit{1}}\ {\isacharparenleft}PCondLeft\ e\ sp\ c\isactrlsub {\isadigit{2}}{\isacharparenright}{\isacharcomma}\ True{\isacharparenright}{\isacharcomma}\ s{\isacharparenright}}}}{[\sc SCondT]}\\[2ex]
\isa{\mbox{}\inferrule{\mbox{eval\ e\ s\ {\isacharequal}\ Bool\ False}}{\mbox{{\isacharparenleft}{\isacharparenleft}Loc\ {\isacharparenleft}Cond\ e\ c\isactrlsub {\isadigit{1}}\ c\isactrlsub {\isadigit{2}}{\isacharparenright}\ sp{\isacharcomma}\ True{\isacharparenright}{\isacharcomma}\ s{\isacharparenright}\ {\isasymrightarrow}\ {\isacharparenleft}{\isacharparenleft}Loc\ c\isactrlsub {\isadigit{2}}\ {\isacharparenleft}PCondRight\ e\ c\isactrlsub {\isadigit{1}}\ sp{\isacharparenright}{\isacharcomma}\ True{\isacharparenright}{\isacharcomma}\ s{\isacharparenright}}}}{[\sc SCondF]}\\[2ex]
\isa{\mbox{}\inferrule{\mbox{eval\ e\ s\ {\isacharequal}\ Bool\ True}}{\mbox{{\isacharparenleft}{\isacharparenleft}Loc\ {\isacharparenleft}While\ e\ c{\isacharparenright}\ sp{\isacharcomma}\ True{\isacharparenright}{\isacharcomma}\ s{\isacharparenright}\ {\isasymrightarrow}\ {\isacharparenleft}{\isacharparenleft}Loc\ c\ {\isacharparenleft}PWhile\ e\ sp{\isacharparenright}{\isacharcomma}\ True{\isacharparenright}{\isacharcomma}\ s{\isacharparenright}}}}{\hspace{1ex} [\sc SWhileT]}\\[2ex]
\isa{\mbox{}\inferrule{\mbox{eval\ e\ s\ {\isacharequal}\ Bool\ False}}{\mbox{{\isacharparenleft}{\isacharparenleft}Loc\ {\isacharparenleft}While\ e\ c{\isacharparenright}\ sp{\isacharcomma}\ True{\isacharparenright}{\isacharcomma}\ s{\isacharparenright}\ {\isasymrightarrow}\ {\isacharparenleft}{\isacharparenleft}Loc\ {\isacharparenleft}While\ e\ c{\isacharparenright}\ sp{\isacharcomma}\ False{\isacharparenright}{\isacharcomma}\ s{\isacharparenright}}}}{\hspace{1ex} [\sc SWhileF]}\\[2ex]
\isa{\mbox{}\inferrule{\mbox{sp\ {\isasymnoteq}\ PTop}}{\mbox{{\isacharparenleft}{\isacharparenleft}Loc\ c\ sp{\isacharcomma}\ False{\isacharparenright}{\isacharcomma}\ s{\isacharparenright}\ {\isasymrightarrow}\ {\isacharparenleft}next{\isacharunderscore}loc\ c\ sp{\isacharcomma}\ s{\isacharparenright}}}}{\hspace{1ex} [\sc SFalse]}
\end{small}
\end{center}%
\end{isamarkuptext}%
\isamarkuptrue%
\figline{}
\caption{Small-step operational semantics}\label{fig:smallStep}
\end{figure}
\begin{isamarkuptext}%
Let us comment on the rules in detail:
\begin{itemize}
\item {\sc SEmpty} executes the \isa{Empty} statement just by swapping the Boolean flag.
\item {\sc SAssign} is similar, but it also updates the state for the assigned variable.
\item {\sc SSeq} moves the pointer to the substatement \isa{c\isactrlsub {\isadigit{1}}}, 
  pushing the substatement \isa{c\isactrlsub {\isadigit{2}}} as continuation to the statement path. 
\item {\sc SCondT} and {\sc SCondF} move to the \emph{then}- respectively \emph{else}- 
  branch of the conditional, depending on the value of the condition. 
\item {\sc SWhileT} moves to the body of the loop.
\item {\sc SWhileF} declares the execution of the loop as terminated, 
  by setting the Boolean flag to \isa{False}.
\item {\sc SFalse} comes into play when execution of the current statement is finished. 
  We then move to the next location, provided we have not already reached the root of 
  the syntax tree and the whole program terminates.
\end{itemize}

The move to the next relevant location is accomplished by function \isa{next{\isacharunderscore}loc} (\figref{fig:next_loc}) which intuitively works as follows: upon conclusion of the first substatement in
a sequence, we move to the second substatement. When finishing the body of a
loop, we move back to the beginning of the loop. In all other cases, we move
up the syntax tree, waiting for rule {\sc SFalse} to relaunch the
function.%
\end{isamarkuptext}%
\isamarkuptrue%
\isamarkupsection{Target language: Automata\label{sec:automata}%
}
\isamarkuptrue%
\isamarkupsubsection{Syntax\label{sec:automata_syntax}%
}
\isamarkuptrue%
\begin{isamarkuptext}%
As usual, our automata are a collection of nodes and edges,
with a distinguished initial state. In this general definition, we will keep the node type \isa{{\isacharprime}n} abstract. It will later be instantiated to \isa{synt{\isacharunderscore}config}.
An edge connects two nodes; moving along an edge may trigger an assignment to a
variable (\isa{AssAct}), or have no effect at all (\isa{NoAct}).

An automaton \isa{{\isacharprime}n\ ta} is a record consisting of a set of \isa{nodes}, a set of \isa{edges} and an initial node \isa{init{\isacharunderscore}s}. An edge has a \isa{source} node, an \isa{action} and a destination node \isa{dest}. Components of a record are written between \isa{{\isasymlparr}\ {\isachardot}{\isachardot}{\isachardot}\ {\isasymrparr}}.%
\end{isamarkuptext}%
\isamarkuptrue%
\isamarkupsubsection{Semantics\label{sec:automata_sem}%
}
\isamarkuptrue%
\begin{isamarkuptext}%
An automaton state is a node, together with a \isa{state} as in \secref{sec:source_sem}.%
\end{isamarkuptext}%
\isamarkuptrue%
\isacommand{type{\isacharunderscore}synonym}\isamarkupfalse%
\ {\isacharprime}n\ ta{\isacharunderscore}state\ {\isacharequal}\ {\isachardoublequoteopen}{\isacharprime}n\ {\isacharasterisk}\ state{\isachardoublequoteclose}%
\begin{isamarkuptext}%
Executing a step of an automaton in an automaton state \isa{{\isacharparenleft}l{\isacharcomma}\ s{\isacharparenright}} consists of selecting an edge starting in node \isa{l}, moving to the
target of the edge and executing its action. Automata are
non-deterministic; in this simplified model, we have no guards for selecting
edges.

\begin{center}
\isa{\mbox{}\inferrule{\mbox{e\ {\isasymin}\ set\ {\isacharparenleft}edges\ aut{\isacharparenright}}\\\ \mbox{l\ {\isacharequal}\ source\ e}\\\ \mbox{l{\isacharprime}\ {\isacharequal}\ dest\ e}\\\ \mbox{s{\isacharprime}\ {\isacharequal}\ action{\isacharunderscore}effect\ {\isacharparenleft}action\ e{\isacharparenright}\ s}}{\mbox{aut\ {\isasymturnstile}\ {\isacharparenleft}l{\isacharcomma}\ s{\isacharparenright}\ {\isasymrightarrow}\ {\isacharparenleft}l{\isacharprime}{\isacharcomma}\ s{\isacharprime}{\isacharparenright}\ }}}{\hspace{1ex} [\sc Action]}
\end{center}%
\end{isamarkuptext}%
\isamarkuptrue%
\isamarkupsection{Automata construction\label{sec:automata_construction}%
}
\isamarkuptrue%
\begin{isamarkuptext}%
The principle of abstracting a statement to an automaton is simple;
the novelty resides in the way the automaton is generated via the zipper
structure: as nodes, we choose the locations of the statements (with their
Boolean flags), and as edges all possible transitions of the semantics.

To make this precise, we need some auxiliary functions. We first define a
function \isa{all{\isacharunderscore}locations} of type \isa{stmt\ {\isasymRightarrow}\ stmt{\isacharunderscore}path\ {\isasymRightarrow}\ stmt{\isacharunderscore}location\ list} which gathers all locations in a statement, and a function \isa{nodes{\isacharunderscore}of{\isacharunderscore}stmt{\isacharunderscore}locations} which adds the Boolean flags.%
\end{isamarkuptext}%
\isamarkuptrue%
\begin{isamarkuptext}%
As for the edges, the function \isa{synt{\isacharunderscore}step{\isacharunderscore}image} yields all
possible successor configurations for a given syntactic configuration. This is of
course an over-approximation of the behavior of the semantics, since some of
the source tree locations may be unreachable during execution.%
\end{isamarkuptext}%
\isamarkuptrue%
\isacommand{fun}\isamarkupfalse%
\ synt{\isacharunderscore}step{\isacharunderscore}image\ {\isacharcolon}{\isacharcolon}\ {\isachardoublequoteopen}synt{\isacharunderscore}config\ {\isasymRightarrow}\ synt{\isacharunderscore}config\ list{\isachardoublequoteclose}\ \isakeyword{where}\isanewline
\ \ {\isachardoublequoteopen}synt{\isacharunderscore}step{\isacharunderscore}image\ {\isacharparenleft}Loc\ Empty\ sp{\isacharcomma}\ True{\isacharparenright}\ {\isacharequal}\ {\isacharbrackleft}{\isacharparenleft}Loc\ Empty\ sp{\isacharcomma}\ False{\isacharparenright}{\isacharbrackright}{\isachardoublequoteclose}\isanewline
{\isacharbar}\ {\isachardoublequoteopen}synt{\isacharunderscore}step{\isacharunderscore}image\ {\isacharparenleft}Loc\ {\isacharparenleft}Assign\ vr\ vl{\isacharparenright}\ sp{\isacharcomma}\ True{\isacharparenright}\ {\isacharequal}\ {\isacharbrackleft}{\isacharparenleft}Loc\ {\isacharparenleft}Assign\ vr\ vl{\isacharparenright}\ sp{\isacharcomma}\ False{\isacharparenright}{\isacharbrackright}{\isachardoublequoteclose}\isanewline
{\isacharbar}\ {\isachardoublequoteopen}synt{\isacharunderscore}step{\isacharunderscore}image\ {\isacharparenleft}Loc\ {\isacharparenleft}Seq\ c{\isadigit{1}}\ c{\isadigit{2}}{\isacharparenright}\ sp{\isacharcomma}\ True{\isacharparenright}\ {\isacharequal}\ {\isacharbrackleft}{\isacharparenleft}Loc\ c{\isadigit{1}}\ {\isacharparenleft}PSeqLeft\ sp\ c{\isadigit{2}}{\isacharparenright}{\isacharcomma}\ True{\isacharparenright}{\isacharbrackright}{\isachardoublequoteclose}\isanewline
{\isacharbar}\ {\isachardoublequoteopen}synt{\isacharunderscore}step{\isacharunderscore}image\ {\isacharparenleft}Loc\ {\isacharparenleft}Cond\ e\ c{\isadigit{1}}\ c{\isadigit{2}}{\isacharparenright}\ sp{\isacharcomma}\ True{\isacharparenright}\ {\isacharequal}\ \isanewline
\ \ \ \ \ \ \ \ \ \ \ \ \ \ {\isacharbrackleft}{\isacharparenleft}Loc\ c{\isadigit{1}}\ {\isacharparenleft}PCondLeft\ e\ sp\ c{\isadigit{2}}{\isacharparenright}{\isacharcomma}\ True{\isacharparenright}{\isacharcomma}\ {\isacharparenleft}Loc\ c{\isadigit{2}}\ {\isacharparenleft}PCondRight\ e\ c{\isadigit{1}}\ sp{\isacharparenright}{\isacharcomma}\ True{\isacharparenright}{\isacharbrackright}{\isachardoublequoteclose}\isanewline
{\isacharbar}\ {\isachardoublequoteopen}synt{\isacharunderscore}step{\isacharunderscore}image\ {\isacharparenleft}Loc\ {\isacharparenleft}While\ e\ c{\isacharparenright}\ sp{\isacharcomma}\ True{\isacharparenright}\ {\isacharequal}\ \isanewline
\ \ \ \ \ \ \ \ \ \ \ \ \ \ {\isacharbrackleft}{\isacharparenleft}Loc\ c\ {\isacharparenleft}PWhile\ e\ sp{\isacharparenright}{\isacharcomma}\ True{\isacharparenright}{\isacharcomma}\ {\isacharparenleft}Loc\ {\isacharparenleft}While\ e\ c{\isacharparenright}\ sp{\isacharcomma}\ False{\isacharparenright}{\isacharbrackright}{\isachardoublequoteclose}\isanewline
{\isacharbar}\ {\isachardoublequoteopen}synt{\isacharunderscore}step{\isacharunderscore}image\ {\isacharparenleft}Loc\ c\ sp{\isacharcomma}\ False{\isacharparenright}\ {\isacharequal}\ {\isacharparenleft}if\ sp\ {\isacharequal}\ PTop\ then\ {\isacharbrackleft}{\isacharbrackright}\ else\ {\isacharbrackleft}next{\isacharunderscore}loc\ c\ sp{\isacharbrackright}{\isacharparenright}{\isachardoublequoteclose}%
\begin{isamarkuptext}%
Together with the following definitions:%
\end{isamarkuptext}%
\isamarkuptrue%
\isacommand{fun}\isamarkupfalse%
\ action{\isacharunderscore}of{\isacharunderscore}synt{\isacharunderscore}config\ {\isacharcolon}{\isacharcolon}\ {\isachardoublequoteopen}synt{\isacharunderscore}config\ {\isasymRightarrow}\ action{\isachardoublequoteclose}\ \isakeyword{where}\isanewline
\ \ {\isachardoublequoteopen}action{\isacharunderscore}of{\isacharunderscore}synt{\isacharunderscore}config\ {\isacharparenleft}Loc\ {\isacharparenleft}Assign\ vn\ vl{\isacharparenright}\ sp{\isacharcomma}\ True{\isacharparenright}\ {\isacharequal}\ AssAct\ vn\ vl{\isachardoublequoteclose}\isanewline
{\isacharbar}\ {\isachardoublequoteopen}action{\isacharunderscore}of{\isacharunderscore}synt{\isacharunderscore}config\ {\isacharparenleft}Loc\ c\ sp{\isacharcomma}\ b{\isacharparenright}\ {\isacharequal}\ NoAct{\isachardoublequoteclose}\isanewline
\isanewline
\isacommand{definition}\isamarkupfalse%
\ edge{\isacharunderscore}of{\isacharunderscore}synt{\isacharunderscore}config\ {\isacharcolon}{\isacharcolon}\ {\isachardoublequoteopen}synt{\isacharunderscore}config\ {\isasymRightarrow}\ synt{\isacharunderscore}config\ edge\ list{\isachardoublequoteclose}\ \isakeyword{where}\isanewline
{\isachardoublequoteopen}edge{\isacharunderscore}of{\isacharunderscore}synt{\isacharunderscore}config\ s\ {\isacharequal}\ \isanewline
map{\isacharparenleft}{\isasymlambda}\ t{\isachardot}\ {\isasymlparr}source\ {\isacharequal}\ s{\isacharcomma}\ action\ {\isacharequal}\ action{\isacharunderscore}of{\isacharunderscore}synt{\isacharunderscore}config\ s{\isacharcomma}\ dest\ {\isacharequal}\ t{\isasymrparr}{\isacharparenright}{\isacharparenleft}synt{\isacharunderscore}step{\isacharunderscore}image\ s{\isacharparenright}{\isachardoublequoteclose}\isanewline
\isacommand{definition}\isamarkupfalse%
\ edges{\isacharunderscore}of{\isacharunderscore}nodes\ {\isacharcolon}{\isacharcolon}\ {\isachardoublequoteopen}synt{\isacharunderscore}config\ list\ {\isasymRightarrow}\ synt{\isacharunderscore}config\ edge\ list{\isachardoublequoteclose}\ \isakeyword{where}\isanewline
\ \ {\isachardoublequoteopen}edges{\isacharunderscore}of{\isacharunderscore}nodes\ nds\ {\isacharequal}\ concat\ {\isacharparenleft}map\ edge{\isacharunderscore}of{\isacharunderscore}synt{\isacharunderscore}config\ nds{\isacharparenright}{\isachardoublequoteclose}%
\begin{isamarkuptext}%
we can define the translation function from statements to automata:%
\end{isamarkuptext}%
\isamarkuptrue%
\isacommand{fun}\isamarkupfalse%
\ stmt{\isacharunderscore}to{\isacharunderscore}ta\ {\isacharcolon}{\isacharcolon}\ {\isachardoublequoteopen}stmt\ {\isasymRightarrow}\ synt{\isacharunderscore}config\ ta{\isachardoublequoteclose}\ \isakeyword{where}\isanewline
\ \ {\isachardoublequoteopen}stmt{\isacharunderscore}to{\isacharunderscore}ta\ c\ {\isacharequal}\ \isanewline
\ \ {\isacharparenleft}let\ nds\ {\isacharequal}\ nodes{\isacharunderscore}of{\isacharunderscore}stmt{\isacharunderscore}locations\ {\isacharparenleft}all{\isacharunderscore}locations\ c\ PTop{\isacharparenright}\ in\isanewline
\ \ \ {\isasymlparr}\ nodes\ {\isacharequal}\ nds{\isacharcomma}\ edges\ {\isacharequal}\ edges{\isacharunderscore}of{\isacharunderscore}nodes\ nds{\isacharcomma}\ init{\isacharunderscore}s\ {\isacharequal}\ {\isacharparenleft}{\isacharparenleft}Loc\ c\ PTop{\isacharparenright}{\isacharcomma}\ True{\isacharparenright}\ {\isasymrparr}{\isacharparenright}{\isachardoublequoteclose}\isanewline
\isadelimproof
\endisadelimproof
\isatagproof
\endisatagproof
{\isafoldproof}%
\isadelimproof
\endisadelimproof
\isadelimproof
\endisadelimproof
\isatagproof
\endisatagproof
{\isafoldproof}%
\isadelimproof
\endisadelimproof
\isadelimproof
\endisadelimproof
\isatagproof
\endisatagproof
{\isafoldproof}%
\isadelimproof
\endisadelimproof
\isadelimproof
\endisadelimproof
\isatagproof
\endisatagproof
{\isafoldproof}%
\isadelimproof
\endisadelimproof
\isadelimproof
\endisadelimproof
\isatagproof
\endisatagproof
{\isafoldproof}%
\isadelimproof
\endisadelimproof
\isadelimproof
\endisadelimproof
\isatagproof
\endisatagproof
{\isafoldproof}%
\isadelimproof
\endisadelimproof
\isadelimproof
\endisadelimproof
\isatagproof
\endisatagproof
{\isafoldproof}%
\isadelimproof
\endisadelimproof
\isadelimproof
\endisadelimproof
\isatagproof
\endisatagproof
{\isafoldproof}%
\isadelimproof
\endisadelimproof
\isadelimproof
\endisadelimproof
\isatagproof
\endisatagproof
{\isafoldproof}%
\isadelimproof
\endisadelimproof
\isadelimproof
\endisadelimproof
\isatagproof
\endisatagproof
{\isafoldproof}%
\isadelimproof
\endisadelimproof
\isadelimproof
\endisadelimproof
\isatagproof
\endisatagproof
{\isafoldproof}%
\isadelimproof
\endisadelimproof
\isamarkupsection{Simulation Property\label{sec:simulation_property}%
}
\isamarkuptrue%
\begin{isamarkuptext}%
We recall that the nodes of the automaton generated by \isa{stmt{\isacharunderscore}to{\isacharunderscore}ta} are labeled by configurations (location, Boolean flag) of the
syntax tree. The simulation lemma (\lemmaref{th:simulation_sem_aut}) holds for
automata with appropriate closure properties: a successor configuration wrt.{}
a transition of the semantics is also a label of the automaton (\isa{nodes{\isacharunderscore}closed}), and analogously for edges (\isa{edges{\isacharunderscore}closed}) or both nodes and edges 
(\isa{synt{\isacharunderscore}step{\isacharunderscore}image{\isacharunderscore}closed}).%
\end{isamarkuptext}%
\isamarkuptrue%
\isadelimproof
\endisadelimproof
\isatagproof
\endisatagproof
{\isafoldproof}%
\isadelimproof
\endisadelimproof
\begin{isamarkuptext}%
The simulation statement is a typical commuting-diagram property: a
step of the program semantics can be simulated by a step of the automaton
semantics, for corresponding program and automata states. For this
correspondence, we use the notation \isa{{\isasymapprox}}, even though it is just plain
syntactic equality in our case.
\begin{lemma}[Simulation property]\label{th:simulation_sem_aut}
\mbox{}\newline
Assume that \isa{synt{\isacharunderscore}step{\isacharunderscore}image{\isacharunderscore}closed\ aut} and \isa{{\isacharparenleft}{\isacharparenleft}{\isacharparenleft}lc{\isacharcomma}\ b{\isacharparenright}{\isacharcomma}\ s{\isacharparenright}\ {\isasymapprox}\ {\isacharparenleft}{\isacharparenleft}lca{\isacharcomma}\ ba{\isacharparenright}{\isacharcomma}\ sa{\isacharparenright}{\isacharparenright}}. If \isa{{\isacharparenleft}{\isacharparenleft}lc{\isacharcomma}\ b{\isacharparenright}{\isacharcomma}\ s{\isacharparenright}\ {\isasymrightarrow}\ {\isacharparenleft}{\isacharparenleft}lc{\isacharprime}{\isacharcomma}\ b{\isacharprime}{\isacharparenright}{\isacharcomma}\ s{\isacharprime}{\isacharparenright}}, then there exist \isa{lca{\isacharprime}{\isacharcomma}\ ba{\isacharprime}{\isacharcomma}\ sa{\isacharprime}} such that \isa{{\isacharparenleft}lca{\isacharprime}{\isacharcomma}\ ba{\isacharprime}{\isacharparenright}\ {\isasymin}\ set\ {\isacharparenleft}nodes\ aut{\isacharparenright}} and the automaton performs the same transition: \isa{aut\ {\isasymturnstile}\ {\isacharparenleft}{\isacharparenleft}lca{\isacharcomma}\ ba{\isacharparenright}{\isacharcomma}\ sa{\isacharparenright}\ {\isasymrightarrow}\ {\isacharparenleft}{\isacharparenleft}lca{\isacharprime}{\isacharcomma}\ ba{\isacharprime}{\isacharparenright}{\isacharcomma}\ sa{\isacharprime}{\isacharparenright}} and \isa{{\isacharparenleft}{\isacharparenleft}lc{\isacharprime}{\isacharcomma}\ b{\isacharprime}{\isacharparenright}{\isacharcomma}\ s{\isacharprime}{\isacharparenright}\ {\isasymapprox}\ {\isacharparenleft}{\isacharparenleft}lca{\isacharprime}{\isacharcomma}\ ba{\isacharprime}{\isacharparenright}{\isacharcomma}\ sa{\isacharprime}{\isacharparenright}}.
\end{lemma}
The proof is a simple induction over the transition relation of the program semantics and is almost fully automatic in the Isabelle proof assistant.%
\end{isamarkuptext}%
\isamarkuptrue%
\isadelimproof
\endisadelimproof
\isatagproof
\endisatagproof
{\isafoldproof}%
\isadelimproof
\endisadelimproof
\isadelimproof
\endisadelimproof
\isatagproof
\endisatagproof
{\isafoldproof}%
\isadelimproof
\endisadelimproof
\isadelimproof
\endisadelimproof
\isatagproof
\endisatagproof
{\isafoldproof}%
\isadelimproof
\endisadelimproof
\isadelimproof
\endisadelimproof
\isatagproof
\endisatagproof
{\isafoldproof}%
\isadelimproof
\endisadelimproof
\begin{isamarkuptext}%
We now want to get rid of the precondition \isa{synt{\isacharunderscore}step{\isacharunderscore}image{\isacharunderscore}closed\ aut} in
\lemmaref{th:simulation_sem_aut}. The first subcase (edge closure), is easy to prove. Node closure is  more difficult and requires the following key lemma:
\begin{lemma}\label{th:synt_step_image_all_locations}
\mbox{}\newline 
\isa{{\normalsize{}If\,}\ lc\ {\isasymin}\ set\ {\isacharparenleft}all{\isacharunderscore}locations\ c\ PTop{\isacharparenright}\ {\normalsize \,then\,}\ set\ {\isacharparenleft}map\ fst\ {\isacharparenleft}synt{\isacharunderscore}step{\isacharunderscore}image\ {\isacharparenleft}lc{\isacharcomma}\ b{\isacharparenright}{\isacharparenright}{\isacharparenright}\ {\isasymsubseteq}\ set\ {\isacharparenleft}all{\isacharunderscore}locations\ c\ PTop{\isacharparenright}{\isachardot}}
\end{lemma}
With this, we obtain the desired
\begin{lemma}[Closure of automaton]\label{th:synt_step_image_closed_stmt_to_ta}
\isa{synt{\isacharunderscore}step{\isacharunderscore}image{\isacharunderscore}closed\ {\isacharparenleft}stmt{\isacharunderscore}to{\isacharunderscore}ta\ c{\isacharparenright}}
\end{lemma}
For the proofs, see \cite{zipper_formalization}.%
\end{isamarkuptext}%
\isamarkuptrue%
\isadelimproof
\endisadelimproof
\isatagproof
\endisatagproof
{\isafoldproof}%
\isadelimproof
\endisadelimproof
\isadelimproof
\endisadelimproof
\isatagproof
\endisatagproof
{\isafoldproof}%
\isadelimproof
\endisadelimproof
\isadelimproof
\endisadelimproof
\isatagproof
\endisatagproof
{\isafoldproof}%
\isadelimproof
\endisadelimproof
\isadelimproof
\endisadelimproof
\isatagproof
\endisatagproof
{\isafoldproof}%
\isadelimproof
\endisadelimproof
\begin{isamarkuptext}%
Let us combine the previous results and write them
more succinctly, by using the notation \isa{{\isasymrightarrow}\isactrlsup {\isacharasterisk}} for the
reflexive-transitive closure for the transition relations of the small-step
semantics and the automaton. Whenever a state is reachable by executing a program \isa{c}
in its initial configuration, then a corresponding (\isa{{\isasymapprox}}) state is reachable 
by running the automaton generated with function \isa{stmt{\isacharunderscore}to{\isacharunderscore}ta}:

\begin{theorem}\label{th:simulation_generated}
\mbox{}\newline 
\isa{{\normalsize{}If\,}\ {\isacharparenleft}{\isacharparenleft}Loc\ c\ PTop{\isacharcomma}\ True{\isacharparenright}{\isacharcomma}\ s{\isacharparenright}\ {\isasymrightarrow}\isactrlsup {\isacharasterisk}\ {\isacharparenleft}cf{\isacharprime}{\isacharcomma}\ s{\isacharprime}{\isacharparenright}\ {\normalsize \,then\,}\ {\isasymexists}cfa{\isacharprime}\ sa{\isacharprime}{\isachardot}\ stmt{\isacharunderscore}to{\isacharunderscore}ta\ c\ {\isasymturnstile}\ {\isacharparenleft}init{\isacharunderscore}s\ {\isacharparenleft}stmt{\isacharunderscore}to{\isacharunderscore}ta\ c{\isacharparenright}{\isacharcomma}\ s{\isacharparenright}\ {\isasymrightarrow}\isactrlsup {\isacharasterisk}\ {\isacharparenleft}cfa{\isacharprime}{\isacharcomma}\ sa{\isacharprime}{\isacharparenright}\ \ {\isasymand}\ {\isacharparenleft}cf{\isacharprime}{\isacharcomma}\ s{\isacharprime}{\isacharparenright}\ {\isasymapprox}\ {\isacharparenleft}cfa{\isacharprime}{\isacharcomma}\ sa{\isacharprime}{\isacharparenright}{\isachardot}}
\end{theorem}

Obviously, the initial configuration of the semantics and the automaton are in
the simulation relation \isa{{\isasymapprox}}, and for the inductive step, we use
\lemmaref{th:simulation_sem_aut}.%
\end{isamarkuptext}%
\isamarkuptrue%
\isadelimtheory
\endisadelimtheory
\isatagtheory
\endisatagtheory
{\isafoldtheory}%
\isadelimtheory
\endisadelimtheory
\end{isabellebody}%

%
\begin{isabellebody}%
\def\isabellecontext{Minimize}%
\isadelimtheory
\endisadelimtheory
\isatagtheory
\endisatagtheory
{\isafoldtheory}%
\isadelimtheory
\endisadelimtheory
\isamarkupsection{Removal of silent transitions\label{sec:silent_transitions}%
}
\isamarkuptrue%
\begin{isamarkuptext}%
Our technique for converting the operational semantics of a program to a finite automaton
generally results in automata containing a large number of silent transitions. Although harmless,
such transitions are only a technical device resulting from the structured nature of operational
semantics: thus, they lack any usefulness in the context of an automaton.

Rather than producing immediately an automaton free of silent transitions, it is possible (and also
quite convenient) to remove them as a final operation. This is obtained by means of a 
\isa{{\isasymtau}}-closure algorithm, where \isa{{\isasymtau}} is the label for silent transitions generally used in
the literature (in our case, \isa{{\isasymtau}\ {\isacharequal}\ NoAct}).

\isa{{\isasymtau}}-closure amounts to computing, for each node in the automaton, the set of those nodes which
can be reached from it by taking any finite number of silent transitions. The following 
\isa{tauclose{\isacharunderscore}step} computes the set of the nodes of an automaton \isa{M} that can be reached 
from a node \isa{s} after taking one silent transition. The argument \isa{x} is used as an 
accumulator when iterating the operation several times, and should be \isa{{\isasymemptyset}} initially:%
\end{isamarkuptext}%
\isamarkuptrue%
\isacommand{definition}\isamarkupfalse%
\ tauclose{\isacharunderscore}step\ {\isacharcolon}{\isacharcolon}\ {\isachardoublequoteopen}{\isacharprime}n\ ta\ {\isasymRightarrow}\ {\isacharprime}n\ {\isasymRightarrow}\ {\isacharprime}n\ set\ {\isasymRightarrow}\ {\isacharprime}n\ set{\isachardoublequoteclose}\ \isakeyword{where}\isanewline
\ \ {\isachardoublequoteopen}tauclose{\isacharunderscore}step\ M\ s\ x\ {\isacharequal}\ {\isacharbraceleft}s{\isacharbraceright}\ {\isasymunion}\ x\ {\isasymunion}\ {\isacharbraceleft}\ n\ {\isasymin}\ set\ {\isacharparenleft}nodes\ M{\isacharparenright}{\isachardot}\ \isanewline
\ \ \ \ \ {\isasymexists}e\ {\isasymin}\ set\ {\isacharparenleft}edges\ M{\isacharparenright}{\isachardot}\ source\ e\ {\isasymin}\ x\ {\isasymand}\ action\ e\ {\isacharequal}\ NoAct\ {\isasymand}\ dest\ e\ {\isacharequal}\ n{\isacharbraceright}{\isachardoublequoteclose}%
\begin{isamarkuptext}%
The proof that \isa{tauclose{\isacharunderscore}step} is monotonically increasing 
(\isa{tauclose{\isacharunderscore}step\ M\ s\ x\ {\isasymsubseteq}\ tauclose{\isacharunderscore}step\ M\ s\ y} for all \isa{x{\isacharcomma}\ y} such that \isa{x\ {\isasymsubseteq}\ y}) is trivial.%
\end{isamarkuptext}%
\isamarkuptrue%
\isacommand{lemma}\isamarkupfalse%
\ mono{\isacharunderscore}tauclose{\isacharunderscore}step\ {\isacharcolon}\ \ {\isachardoublequoteopen}mono\ {\isacharparenleft}tauclose{\isacharunderscore}step\ M\ s{\isacharparenright}{\isachardoublequoteclose}%
\isadelimproof
\endisadelimproof
\isatagproof
\endisatagproof
{\isafoldproof}%
\isadelimproof
\endisadelimproof
\begin{isamarkuptext}%
Then, the operation \isa{tauclose} is defined as the least fixpoint of the monotonic
operator:%
\end{isamarkuptext}%
\isamarkuptrue%
\isacommand{definition}\isamarkupfalse%
\ tauclose\ {\isacharcolon}{\isacharcolon}\ {\isachardoublequoteopen}{\isacharprime}n\ ta\ {\isasymRightarrow}\ {\isacharprime}n\ {\isasymRightarrow}\ {\isacharprime}n\ set{\isachardoublequoteclose}\ \isakeyword{where}\isanewline
{\isachardoublequoteopen}tauclose\ M\ n\ {\isacharequal}\ lfp\ {\isacharparenleft}tauclose{\isacharunderscore}step\ M\ n{\isacharparenright}{\isachardoublequoteclose}%
\begin{isamarkuptext}%
To obtain a \isa{{\isasymtau}}-closed automaton, we simply map the nodes of the input automaton to their 
\isa{{\isasymtau}}-closed counterpart (and similarly for the initial node). To compute the set of edges, we
consider the rationale behind the definition of the \isa{{\isasymtau}}-closure of an automaton. Informally,
being in a certain node or in any other node reachable from it only by means of silent 
transitions, is equivalent. When we compute the \isa{{\isasymtau}}-closure of a certain node, we are 
essentially identifying all the nodes in it: thus the edges with source \isa{tauclose\ M\ s{\isadigit{1}}}
should be those that leave any of the nodes in the \isa{{\isasymtau}}-closure. To make things more formal, 
let us introduce the notation $x \stackrel{\alpha}{\longrightarrow} y$ for edges going from node $x$ to
node $y$ labeled with action $\alpha$: using this notation, the edges of 
the \isa{{\isasymtau}}-closed automaton are taken to be those in the form \isa{tauclose\ M\ s{\isadigit{1}}} 
$\stackrel{\alpha}{\longrightarrow}$ \isa{tauclose\ M\ s{\isadigit{2}}}, such that for some \isa{s\ {\isasymin}\ tauclose\ M\ s{\isadigit{1}}}, \isa{s} $\stackrel{\alpha}{\longrightarrow}$ \isa{s{\isadigit{2}}} is a non-silent 
transition in the input automaton. 

\begin{figure}
{\fontsize{2.8mm}{1em}\selectfont
\begin{center}
\begin{minipage}{0.45\textwidth}
\begin{center}
\begin{tikzcd}
\phantom{.} & 1 \arrow{dl}{\tau} \arrow{dr}{\tau} & \\
2 \arrow{d}{\alpha} & & 3 \arrow[bend left=20]{d}{\beta}  \\
4 \arrow{rr}{\beta} & & 5 \arrow[bend left=20]{u}{\gamma}
\end{tikzcd}
\end{center}
\end{minipage}
\begin{minipage}{0.45\textwidth}
\begin{center}
\begin{tikzcd}
\phantom{.} & \{1,2,3\} \arrow{ddl}{\alpha} \arrow{ddr}{\beta} & \\
\{2\} \arrow{d}{\alpha} & & \{3\} \arrow[bend left=20]{d}{\beta}  \\
\{4\} \arrow{rr}{\beta} & & \{5\} \arrow[bend left=20]{u}{\gamma}
\end{tikzcd}
\end{center}
\end{minipage}
\end{center}
}
\caption{A simple automaton and its $\tau$-closure.}
\label{tauclosure}
\end{figure}
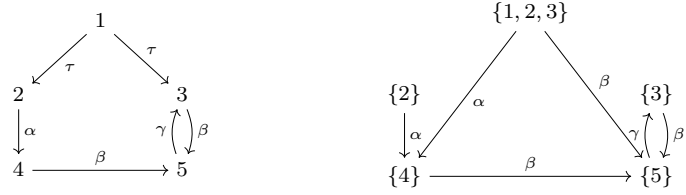%
\end{isamarkuptext}%
\isamarkuptrue%
\isacommand{definition}\isamarkupfalse%
\ tauclose{\isacharunderscore}nodes\ {\isacharcolon}{\isacharcolon}\ {\isachardoublequoteopen}{\isacharprime}n\ ta\ {\isasymRightarrow}\ {\isacharprime}n\ set\ list{\isachardoublequoteclose}\ \isakeyword{where}\isanewline
\ \ {\isachardoublequoteopen}tauclose{\isacharunderscore}nodes\ M\ {\isacharequal}\ List{\isachardot}map\ {\isacharparenleft}tauclose\ M{\isacharparenright}\ {\isacharparenleft}nodes\ M{\isacharparenright}{\isachardoublequoteclose}\isanewline
\isacommand{definition}\isamarkupfalse%
\ tauclose{\isacharunderscore}init{\isacharunderscore}s\ {\isacharcolon}{\isacharcolon}\ {\isachardoublequoteopen}{\isacharprime}n\ ta\ {\isasymRightarrow}\ {\isacharprime}n\ set{\isachardoublequoteclose}\ \isakeyword{where}\isanewline
\ \ {\isachardoublequoteopen}tauclose{\isacharunderscore}init{\isacharunderscore}s\ M\ {\isacharequal}\ tauclose\ M\ {\isacharparenleft}init{\isacharunderscore}s\ M{\isacharparenright}{\isachardoublequoteclose}\isanewline
\isanewline
\isacommand{definition}\isamarkupfalse%
\ acts{\isacharunderscore}of{\isacharunderscore}ta\ {\isacharcolon}{\isacharcolon}\ {\isachardoublequoteopen}{\isacharprime}n\ ta\ {\isasymRightarrow}\ action\ list{\isachardoublequoteclose}\ \isakeyword{where}\isanewline
\ \ {\isachardoublequoteopen}acts{\isacharunderscore}of{\isacharunderscore}ta\ M\ {\isacharequal}\ List{\isachardot}map\ {\isacharparenleft}{\isasymlambda}e{\isachardot}{\isacharparenleft}action\ e{\isacharparenright}{\isacharparenright}\ {\isacharparenleft}edges\ M{\isacharparenright}{\isachardoublequoteclose}\isanewline
\isanewline
\isacommand{definition}\isamarkupfalse%
\ possible{\isacharunderscore}tau{\isacharunderscore}edges\ {\isacharcolon}{\isacharcolon}\ {\isachardoublequoteopen}{\isacharprime}n\ ta\ {\isasymRightarrow}\ {\isacharprime}n\ set\ edge\ list{\isachardoublequoteclose}\ \isakeyword{where}\isanewline
\ \ {\isachardoublequoteopen}possible{\isacharunderscore}tau{\isacharunderscore}edges\ M\ {\isacharequal}\ \isanewline
\ \ \ \ List{\isachardot}map\ {\isacharparenleft}{\isasymlambda}{\isacharparenleft}s{\isacharcomma}a{\isacharcomma}t{\isacharparenright}{\isachardot}{\isasymlparr}source\ {\isacharequal}\ tauclose\ M\ s{\isacharcomma}action\ {\isacharequal}\ a{\isacharcomma}dest\ {\isacharequal}\ tauclose\ M\ t{\isasymrparr}{\isacharparenright}\ \isanewline
\ \ \ \ \ {\isacharparenleft}List{\isachardot}product\ {\isacharparenleft}nodes\ M{\isacharparenright}\ {\isacharparenleft}List{\isachardot}product\ {\isacharparenleft}acts{\isacharunderscore}of{\isacharunderscore}ta\ M{\isacharparenright}\ {\isacharparenleft}nodes\ M{\isacharparenright}{\isacharparenright}{\isacharparenright}{\isachardoublequoteclose}\isanewline
\isanewline
\isacommand{definition}\isamarkupfalse%
\ tauclose{\isacharunderscore}edges\ {\isacharcolon}{\isacharcolon}\ {\isachardoublequoteopen}{\isacharprime}n\ ta\ {\isasymRightarrow}\ {\isacharprime}n\ set\ edge\ list{\isachardoublequoteclose}\ \isakeyword{where}\isanewline
\ \ {\isachardoublequoteopen}tauclose{\isacharunderscore}edges\ M\ {\isacharequal}\ List{\isachardot}filter\ \isanewline
\ \ \ \ {\isacharparenleft}{\isasymlambda}e{\isachardot}{\isacharparenleft}{\isasymexists}s{\isadigit{1}}\ a\ s{\isadigit{2}}{\isachardot}{\isacharparenleft}e\ {\isacharequal}\ {\isasymlparr}source\ {\isacharequal}\ tauclose\ M\ s{\isadigit{1}}{\isacharcomma}action\ {\isacharequal}\ a{\isacharcomma}dest\ {\isacharequal}\ tauclose\ M\ s{\isadigit{2}}{\isasymrparr}\ {\isasymand}\ \isanewline
\ \ \ \ \ \ \ a\ {\isasymnoteq}\ NoAct\ {\isasymand}\ \isanewline
\ \ \ \ \ \ \ {\isacharparenleft}{\isasymexists}s\ {\isasymin}\ tauclose\ M\ s{\isadigit{1}}{\isachardot}{\isasymlparr}source\ {\isacharequal}\ s{\isacharcomma}action\ {\isacharequal}\ a{\isacharcomma}dest\ {\isacharequal}\ s{\isadigit{2}}{\isasymrparr}\ {\isasymin}\ set\ {\isacharparenleft}edges\ M{\isacharparenright}{\isacharparenright}{\isacharparenright}{\isacharparenright}{\isacharparenright}\isanewline
\ \ \ \ {\isacharparenleft}possible{\isacharunderscore}tau{\isacharunderscore}edges\ M{\isacharparenright}{\isachardoublequoteclose}\isanewline
\isanewline
\isacommand{definition}\isamarkupfalse%
\ tauclose{\isacharunderscore}ta\ {\isacharcolon}{\isacharcolon}\ {\isachardoublequoteopen}{\isacharprime}n\ ta\ {\isasymRightarrow}\ {\isacharprime}n\ set\ ta{\isachardoublequoteclose}\ \isakeyword{where}\isanewline
\ \ {\isachardoublequoteopen}tauclose{\isacharunderscore}ta\ M\ {\isacharequal}\ {\isasymlparr}nodes\ {\isacharequal}\ tauclose{\isacharunderscore}nodes\ M{\isacharcomma}\ \isanewline
\ \ \ \ \ \ \ \ \ \ \ \ \ \ \ \ \ \ \ \ edges\ {\isacharequal}\ tauclose{\isacharunderscore}edges\ M{\isacharcomma}\ \isanewline
\ \ \ \ \ \ \ \ \ \ \ \ \ \ \ \ \ \ \ \ init{\isacharunderscore}s\ {\isacharequal}\ tauclose{\isacharunderscore}init{\isacharunderscore}s\ M\ {\isasymrparr}{\isachardoublequoteclose}%
\begin{isamarkuptext}%
The automaton obtained by \isa{{\isasymtau}}-closure
(see example in Figure~\ref{tauclosure}) has no silent edges any more: when a
silent transition is taken in the input automaton, the corresponding operation in its 
\isa{{\isasymtau}}-closure is to stay in the same node; when a non-silent transition 
$s \stackrel{\alpha}{\longrightarrow} s'$ is taken in the input automaton, a transition with the 
same label and target is taken in its \isa{{\isasymtau}}-closure: however the source of this transition does 
not have to be \isa{tauclose\ s}, but can be the \isa{{\isasymtau}}-closure of any node from which $s$ can be 
reached by taking silent transitions.

This correspondence between an automaton and its \isa{{\isasymtau}}-closure, is expressed by the following 
simulation:%
\end{isamarkuptext}%
\isamarkuptrue%
\isacommand{definition}\isamarkupfalse%
\ tau{\isacharunderscore}sim\ {\isacharcolon}{\isacharcolon}\ {\isachardoublequoteopen}{\isacharprime}n{\isadigit{1}}\ ta\ {\isasymRightarrow}\ {\isacharprime}n{\isadigit{2}}\ ta\ {\isasymRightarrow}\ bool{\isachardoublequoteclose}\ \isakeyword{where}\isanewline
{\isachardoublequoteopen}tau{\isacharunderscore}sim\ M{\isadigit{1}}\ M{\isadigit{2}}\ {\isacharequal}\isanewline
\ \ {\isacharparenleft}{\isasymexists}R{\isachardot}\ R\ {\isacharparenleft}init{\isacharunderscore}s\ M{\isadigit{1}}{\isacharparenright}\ {\isacharparenleft}init{\isacharunderscore}s\ M{\isadigit{2}}{\isacharparenright}\ \ {\isasymand}\isanewline
\ \ \ \ \ \ {\isacharparenleft}{\isasymforall}s{\isadigit{1}}\ s{\isadigit{2}}{\isachardot}\ R\ s{\isadigit{1}}\ s{\isadigit{2}}\ {\isasymlongrightarrow}\isanewline
\ \ \ \ \ \ \ \ {\isacharparenleft}{\isasymforall}s{\isadigit{1}}{\isacharprime}\ a{\isachardot}\ {\isasymlparr}source\ {\isacharequal}\ s{\isadigit{1}}{\isacharcomma}action\ {\isacharequal}\ a{\isacharcomma}dest\ {\isacharequal}\ s{\isadigit{1}}{\isacharprime}{\isasymrparr}\ {\isasymin}\ set\ {\isacharparenleft}edges\ M{\isadigit{1}}{\isacharparenright}\ {\isasymlongrightarrow}\isanewline
\ \ \ \ \ \ \ \ \ \ \ {\isacharparenleft}a\ {\isacharequal}\ NoAct\ {\isasymand}\ R\ s{\isadigit{1}}{\isacharprime}\ s{\isadigit{2}}{\isacharparenright}\ {\isasymor}\ \isanewline
\ \ \ \ \ \ \ \ \ \ \ {\isacharparenleft}{\isasymexists}s{\isadigit{2}}{\isacharprime}{\isachardot}{\isasymlparr}source\ {\isacharequal}\ s{\isadigit{2}}{\isacharcomma}action\ {\isacharequal}\ a{\isacharcomma}dest\ {\isacharequal}\ s{\isadigit{2}}{\isacharprime}{\isasymrparr}\ {\isasymin}\ set\ {\isacharparenleft}edges\ M{\isadigit{2}}{\isacharparenright}\ {\isasymand}\ R\ s{\isadigit{1}}{\isacharprime}\ s{\isadigit{2}}{\isacharprime}{\isacharparenright}{\isacharparenright}{\isacharparenright}{\isacharparenright}{\isachardoublequoteclose}%
\begin{isamarkuptext}%
In our case, we shall instantiate the type parameter \isa{{\isacharprime}n{\isadigit{2}}}  with \isa{{\isacharprime}n{\isadigit{1}}\ set} and take the relation \isa{R} to be such that 
\isa{R\ s\ s{\isacharprime}\ {\isasymLongleftrightarrow}\ {\isacharparenleft}s\ {\isasymin}\ set\ {\isacharparenleft}nodes\ M{\isacharparenright}\ {\isasymand}\ s{\isacharprime}\ {\isasymin}\ set\ {\isacharparenleft}nodes\ {\isacharparenleft}tauclose{\isacharunderscore}ta\ M{\isacharparenright}{\isacharparenright}\ {\isasymand}\ s\ {\isasymin}\ s{\isacharprime}{\isacharparenright}}.%
\end{isamarkuptext}%
\isamarkuptrue%
\isadelimproof
\endisadelimproof
\isatagproof
\endisatagproof
{\isafoldproof}%
\isadelimproof
\endisadelimproof
\isadelimproof
\endisadelimproof
\isatagproof
\endisatagproof
{\isafoldproof}%
\isadelimproof
\endisadelimproof
\begin{isamarkuptext}%
We are able to prove the simulation for all well formed automata. An automaton is
well formed (\isa{regular{\isacharunderscore}ta}) when its initial nodes and the sources and targets of all its edges 
are in the set of its nodes.%
\end{isamarkuptext}%
\isamarkuptrue%
\isacommand{definition}\isamarkupfalse%
\ regular{\isacharunderscore}ta\ {\isacharcolon}{\isacharcolon}\ {\isachardoublequoteopen}{\isacharprime}n\ ta\ {\isasymRightarrow}\ bool{\isachardoublequoteclose}\ \isakeyword{where}\ \isanewline
\ \ {\isachardoublequoteopen}regular{\isacharunderscore}ta\ M\ {\isacharequal}\ \isanewline
\ \ \ \ {\isacharparenleft}init{\isacharunderscore}s\ M\ {\isasymin}\ set\ {\isacharparenleft}nodes\ M{\isacharparenright}\ {\isasymand}\ \isanewline
\ \ \ \ \ \ {\isacharparenleft}{\isasymforall}e\ {\isasymin}\ set\ {\isacharparenleft}edges\ M{\isacharparenright}{\isachardot}\ source\ e\ {\isasymin}\ set\ {\isacharparenleft}nodes\ M{\isacharparenright}\ {\isasymand}\ dest\ e\ {\isasymin}\ set\ {\isacharparenleft}nodes\ M{\isacharparenright}{\isacharparenright}{\isacharparenright}{\isachardoublequoteclose}%
\isadelimproof
\endisadelimproof
\isatagproof
\endisatagproof
{\isafoldproof}%
\isadelimproof
\endisadelimproof
\isadelimproof
\endisadelimproof
\isatagproof
\endisatagproof
{\isafoldproof}%
\isadelimproof
\endisadelimproof
\isadelimproof
\endisadelimproof
\isatagproof
\endisatagproof
{\isafoldproof}%
\isadelimproof
\endisadelimproof
\isadelimproof
\endisadelimproof
\isatagproof
\endisatagproof
{\isafoldproof}%
\isadelimproof
\endisadelimproof
\isadelimproof
\endisadelimproof
\isatagproof
\endisatagproof
{\isafoldproof}%
\isadelimproof
\endisadelimproof
\isadelimproof
\endisadelimproof
\isatagproof
\endisatagproof
{\isafoldproof}%
\isadelimproof
\endisadelimproof
\isadelimproof
\endisadelimproof
\isatagproof
\endisatagproof
{\isafoldproof}%
\isadelimproof
\endisadelimproof
\isadelimproof
\endisadelimproof
\isatagproof
\endisatagproof
{\isafoldproof}%
\isadelimproof
\endisadelimproof
\isadelimproof
\endisadelimproof
\isatagproof
\endisatagproof
{\isafoldproof}%
\isadelimproof
\endisadelimproof
\isadelimproof
\endisadelimproof
\isatagproof
\endisatagproof
{\isafoldproof}%
\isadelimproof
\endisadelimproof
\isadelimproof
\endisadelimproof
\isatagproof
\endisatagproof
{\isafoldproof}%
\isadelimproof
\endisadelimproof
\isadelimproof
\endisadelimproof
\isatagproof
\endisatagproof
{\isafoldproof}%
\isadelimproof
\endisadelimproof
\begin{isamarkuptext}%
\begin{theorem}[simulation of $\tau$-closure]\label{th:sim_tauclose_ta}
\mbox{}\newline
\isa{{\normalsize{}If\,}\ regular{\isacharunderscore}ta\ M\ {\normalsize \,then\,}\ tau{\isacharunderscore}sim\ M\ {\isacharparenleft}tauclose{\isacharunderscore}ta\ M{\isacharparenright}{\isachardot}}
\end{theorem}

The proof follows from the definitions, proceeding by cases on the possible actions.%
\end{isamarkuptext}%
\isamarkuptrue%
\isadelimproof
\endisadelimproof
\isatagproof
\endisatagproof
{\isafoldproof}%
\isadelimproof
\endisadelimproof
\isadelimproof
\endisadelimproof
\isatagproof
\endisatagproof
{\isafoldproof}%
\isadelimproof
\endisadelimproof
\begin{isamarkuptext}%
As a final remark, it is worth noting that the definition of \isa{tauclose} is not 
entirely satisfying, given that there exists no general method to compute a fixpoint in a finite
amount of time. In our case, however, the fixpoint can be computed by iterating the \isa{tauclose{\isacharunderscore}step} function, since it is monotonically increasing with a finite upper bound, namely 
the set of nodes of the input automaton. Thus, we can define the following ``computational'' version 
of the \isa{{\isasymtau}}-closure operation:%
\end{isamarkuptext}%
\isamarkuptrue%
\isacommand{function}\isamarkupfalse%
\ tauclose{\isacharunderscore}comp{\isacharunderscore}aux\ {\isacharcolon}{\isacharcolon}\ {\isachardoublequoteopen}{\isacharprime}n\ ta\ {\isasymRightarrow}\ {\isacharprime}n\ {\isasymRightarrow}\ {\isacharprime}n\ set\ {\isasymRightarrow}\ {\isacharprime}n\ set{\isachardoublequoteclose}\ \isakeyword{where}\isanewline
\ \ {\isachardoublequoteopen}tauclose{\isacharunderscore}step\ M\ s\ x\ {\isacharequal}\ x\ {\isasymLongrightarrow}\ \isanewline
\ \ \ \ \ tauclose{\isacharunderscore}comp{\isacharunderscore}aux\ M\ s\ x\ {\isacharequal}\ x{\isachardoublequoteclose}\isanewline
{\isacharbar}\ {\isachardoublequoteopen}tauclose{\isacharunderscore}step\ M\ s\ x\ {\isasymnoteq}\ x\ {\isasymLongrightarrow}\ \isanewline
\ \ \ \ \ tauclose{\isacharunderscore}comp{\isacharunderscore}aux\ M\ s\ x\ {\isacharequal}\ tauclose{\isacharunderscore}comp{\isacharunderscore}aux\ M\ s\ {\isacharparenleft}tauclose{\isacharunderscore}step\ M\ s\ x{\isacharparenright}{\isachardoublequoteclose}\isanewline
\isadelimproof
\endisadelimproof
\isatagproof
\isacommand{by}\isamarkupfalse%
\ {\isacharparenleft}atomize{\isacharunderscore}elim{\isacharcomma}auto{\isacharparenright}%
\endisatagproof
{\isafoldproof}%
\isadelimproof
\endisadelimproof
\begin{isamarkuptext}%
\noindent \isa{{\isacharparenleft}{\isacharasterisk}\ termination\ proof\ omitted\ {\isacharasterisk}{\isacharparenright}}%
\end{isamarkuptext}%
\isamarkuptrue%
\isacommand{termination}\isamarkupfalse%
\isadelimproof
\ %
\endisadelimproof
\isatagproof
\isacommand{proof}\isamarkupfalse%
\isanewline
\ \ {\isacharparenleft}relation\ {\isachardoublequoteopen}measure\ {\isacharparenleft}{\isasymlambda}{\isacharparenleft}M{\isacharcomma}s{\isacharcomma}x{\isacharparenright}{\isachardot}length\ {\isacharparenleft}filter\ {\isacharparenleft}{\isasymlambda}v{\isachardot}{\isacharparenleft}v\ {\isasymnotin}\ x{\isacharparenright}{\isacharparenright}\ {\isacharparenleft}s\ {\isacharhash}\ nodes\ M{\isacharparenright}{\isacharparenright}{\isacharparenright}{\isachardoublequoteclose}{\isacharcomma}\isanewline
\ \ \ simp{\isacharcomma}unfold\ measure{\isacharunderscore}def{\isacharparenright}\isanewline
\ \ \isacommand{fix}\isamarkupfalse%
\ M\ s\ x\isanewline
\ \ \isacommand{assume}\isamarkupfalse%
\ hneq{\isacharcolon}{\isachardoublequoteopen}tauclose{\isacharunderscore}step\ M\ s\ x\ {\isasymnoteq}\ x{\isachardoublequoteclose}\isanewline
\ \ \isacommand{from}\isamarkupfalse%
\ hneq\ mono{\isacharunderscore}tauclose{\isacharunderscore}step\ \isacommand{have}\isamarkupfalse%
\ {\isachardoublequoteopen}{\isasymexists}c{\isachardot}{\isacharparenleft}c\ {\isasymin}\ tauclose{\isacharunderscore}step\ M\ s\ x\ {\isasymand}\ c\ {\isasymnotin}\ x{\isacharparenright}{\isachardoublequoteclose}\isanewline
\ \ \ \ \isacommand{by}\isamarkupfalse%
\ {\isacharparenleft}unfold\ mono{\isacharunderscore}def\ tauclose{\isacharunderscore}step{\isacharunderscore}def{\isacharcomma}auto{\isacharparenright}\isanewline
\ \ \isacommand{from}\isamarkupfalse%
\ this\ \isacommand{obtain}\isamarkupfalse%
\ c\ \isakeyword{where}\ hcin{\isacharcolon}{\isachardoublequoteopen}c\ {\isasymin}\ tauclose{\isacharunderscore}step\ M\ s\ x{\isachardoublequoteclose}\ \isakeyword{and}\ hcnotin{\isacharcolon}{\isachardoublequoteopen}c\ {\isasymnotin}\ x{\isachardoublequoteclose}\ \isacommand{by}\isamarkupfalse%
\ blast\isanewline
\ \ \isacommand{have}\isamarkupfalse%
\ hmagic{\isacharcolon}\isanewline
\ \ \ \ {\isachardoublequoteopen}length\ {\isacharbrackleft}v{\isasymleftarrow}s\ {\isacharhash}\ nodes\ M\ {\isachardot}\ v\ {\isasymnotin}\ tauclose{\isacharunderscore}step\ M\ s\ x{\isacharbrackright}\ \isanewline
\ \ \ \ \ \ \ {\isacharless}\ length\ {\isacharbrackleft}v{\isasymleftarrow}\ s\ {\isacharhash}\ nodes\ M\ {\isachardot}\ v\ {\isasymnotin}\ x{\isacharbrackright}\ {\isasymLongrightarrow}\isanewline
\ \ \ \ \ {\isacharparenleft}{\isacharparenleft}M{\isacharcomma}\ s{\isacharcomma}\ tauclose{\isacharunderscore}step\ M\ s\ x{\isacharparenright}{\isacharcomma}\ M{\isacharcomma}\ s{\isacharcomma}\ x{\isacharparenright}\ \isanewline
\ \ \ \ \ \ \ {\isasymin}\ inv{\isacharunderscore}image\ less{\isacharunderscore}than\ {\isacharparenleft}{\isasymlambda}{\isacharparenleft}M{\isacharcomma}\ s{\isacharcomma}\ x{\isacharparenright}{\isachardot}\ length\ {\isacharbrackleft}v{\isasymleftarrow}s\ {\isacharhash}\ nodes\ M\ {\isachardot}\ v\ {\isasymnotin}\ x{\isacharbrackright}{\isacharparenright}{\isachardoublequoteclose}\isanewline
\ \ \ \ \isacommand{by}\isamarkupfalse%
\ {\isacharparenleft}simp{\isacharparenright}\isanewline
\ \ \isacommand{from}\isamarkupfalse%
\ hneq\ \isacommand{have}\isamarkupfalse%
\ {\isachardoublequoteopen}x\ {\isasymsubset}\ tauclose{\isacharunderscore}step\ M\ s\ x{\isachardoublequoteclose}\ \isacommand{by}\isamarkupfalse%
\ {\isacharparenleft}unfold\ tauclose{\isacharunderscore}step{\isacharunderscore}def{\isacharcomma}auto{\isacharparenright}\isanewline
\ \ \isacommand{moreover}\isamarkupfalse%
\ \isacommand{from}\isamarkupfalse%
\ hcin\ hcnotin\ \isacommand{have}\isamarkupfalse%
\ {\isachardoublequoteopen}c\ {\isasymin}\ set\ {\isacharparenleft}s\ {\isacharhash}\ nodes\ M{\isacharparenright}{\isachardoublequoteclose}\ \isacommand{by}\isamarkupfalse%
\ {\isacharparenleft}unfold\ tauclose{\isacharunderscore}step{\isacharunderscore}def{\isacharcomma}auto{\isacharparenright}\isanewline
\ \ \isacommand{moreover}\isamarkupfalse%
\ \isacommand{note}\isamarkupfalse%
\ hcin\ hcnotin\isanewline
\ \ \isacommand{ultimately}\isamarkupfalse%
\ \isacommand{have}\isamarkupfalse%
\ \isanewline
\ \ \ \ {\isachardoublequoteopen}length\ {\isacharbrackleft}v{\isasymleftarrow}s\ {\isacharhash}\ nodes\ M\ {\isachardot}\ v\ {\isasymnotin}\ tauclose{\isacharunderscore}step\ M\ s\ x{\isacharbrackright}\isanewline
\ \ \ \ \ \ \ {\isacharless}\ length\ {\isacharbrackleft}v{\isasymleftarrow}\ s\ {\isacharhash}\ nodes\ M\ {\isachardot}\ v\ {\isasymnotin}\ x{\isacharbrackright}{\isachardoublequoteclose}\isanewline
\ \ \ \ \isacommand{by}\isamarkupfalse%
\ {\isacharparenleft}rule{\isacharunderscore}tac\ filter{\isacharunderscore}subset{\isacharcomma}auto{\isacharparenright}\isanewline
\ \ \isacommand{from}\isamarkupfalse%
\ this\ hmagic\ \isacommand{show}\isamarkupfalse%
\ \isanewline
\ \ \ \ {\isachardoublequoteopen}{\isacharparenleft}{\isacharparenleft}M{\isacharcomma}\ s{\isacharcomma}\ tauclose{\isacharunderscore}step\ M\ s\ x{\isacharparenright}{\isacharcomma}\ M{\isacharcomma}\ s{\isacharcomma}\ x{\isacharparenright}\ \isanewline
\ \ \ \ \ \ {\isasymin}\ inv{\isacharunderscore}image\ less{\isacharunderscore}than\ {\isacharparenleft}{\isasymlambda}{\isacharparenleft}M{\isacharcomma}\ s{\isacharcomma}\ x{\isacharparenright}{\isachardot}\ length\ {\isacharbrackleft}v{\isasymleftarrow}s\ {\isacharhash}\ nodes\ M\ {\isachardot}\ v\ {\isasymnotin}\ x{\isacharbrackright}{\isacharparenright}{\isachardoublequoteclose}\ \isanewline
\ \ \ \ \isacommand{by}\isamarkupfalse%
\ auto\isanewline
\isacommand{qed}\isamarkupfalse%
\endisatagproof
{\isafoldproof}%
\isadelimproof
\endisadelimproof
\isanewline
\isanewline
\isacommand{definition}\isamarkupfalse%
\ tauclose{\isacharunderscore}comp\ {\isacharcolon}{\isacharcolon}\ {\isachardoublequoteopen}{\isacharprime}n\ ta\ {\isasymRightarrow}\ {\isacharprime}n\ {\isasymRightarrow}\ {\isacharprime}n\ set{\isachardoublequoteclose}\ \isakeyword{where}\isanewline
{\isachardoublequoteopen}tauclose{\isacharunderscore}comp\ M\ s\ {\isacharequal}\ tauclose{\isacharunderscore}comp{\isacharunderscore}aux\ M\ s\ {\isacharbraceleft}{\isacharbraceright}{\isachardoublequoteclose}%
\begin{isamarkuptext}%
The function \isa{tauclose{\isacharunderscore}comp{\isacharunderscore}aux} cannot be proved to be total automatically: we provide such 
a proof based on the finite upper bound argument we have just mentioned. As expected, we can show 
that \isa{tauclose} and \isa{tauclose{\isacharunderscore}comp} compute the same function.%
\end{isamarkuptext}%
\isamarkuptrue%
\isacommand{lemma}\isamarkupfalse%
\ tauclose{\isacharunderscore}comp{\isacharunderscore}aux{\isacharunderscore}sound\ {\isacharcolon}\isanewline
\ \ \isakeyword{assumes}\ {\isachardoublequoteopen}x\ {\isasymsubseteq}\ tauclose\ M\ s{\isachardoublequoteclose}\isanewline
\ \ \isakeyword{shows}\ {\isachardoublequoteopen}tauclose{\isacharunderscore}comp{\isacharunderscore}aux\ M\ s\ x\ {\isacharequal}\ tauclose\ M\ s{\isachardoublequoteclose}\isanewline
\isadelimproof
\endisadelimproof
\isatagproof
\isacommand{using}\isamarkupfalse%
\ assms\isanewline
\isacommand{proof}\isamarkupfalse%
\ {\isacharparenleft}induct\ M\ s\ x\ rule{\isacharcolon}tauclose{\isacharunderscore}comp{\isacharunderscore}aux{\isachardot}induct{\isacharcomma}unfold\ tauclose{\isacharunderscore}def{\isacharcomma}simp{\isacharparenright}\isanewline
\ \ \isacommand{fix}\isamarkupfalse%
\ Ma\ sa\ xa\isanewline
\ \ \isacommand{assume}\isamarkupfalse%
\ {\isachardoublequoteopen}tauclose{\isacharunderscore}step\ Ma\ sa\ xa\ {\isacharequal}\ xa{\isachardoublequoteclose}\ {\isachardoublequoteopen}xa\ {\isasymsubseteq}\ lfp\ {\isacharparenleft}tauclose{\isacharunderscore}step\ Ma\ sa{\isacharparenright}{\isachardoublequoteclose}\isanewline
\ \ \isacommand{from}\isamarkupfalse%
\ this\ \isacommand{show}\isamarkupfalse%
\ {\isachardoublequoteopen}xa\ {\isacharequal}\ lfp\ {\isacharparenleft}tauclose{\isacharunderscore}step\ Ma\ sa{\isacharparenright}{\isachardoublequoteclose}\ \isacommand{by}\isamarkupfalse%
\ {\isacharparenleft}unfold\ lfp{\isacharunderscore}def{\isacharcomma}auto{\isacharparenright}\isanewline
\isacommand{next}\isamarkupfalse%
\isanewline
\ \ \isacommand{fix}\isamarkupfalse%
\ Ma\ sa\ xa\isanewline
\ \ \isacommand{assume}\isamarkupfalse%
\ \ \ \ {\isachardoublequoteopen}tauclose{\isacharunderscore}step\ Ma\ sa\ xa\ {\isasymnoteq}\ xa{\isachardoublequoteclose}\isanewline
\ \ \ \ \ \isakeyword{and}\ ih{\isacharcolon}{\isachardoublequoteopen}tauclose{\isacharunderscore}step\ Ma\ sa\ xa\ {\isasymsubseteq}\ lfp\ {\isacharparenleft}tauclose{\isacharunderscore}step\ Ma\ sa{\isacharparenright}\ {\isasymLongrightarrow}\isanewline
\ \ \ \ \ \ \ \ \ \ \ \ \ tauclose{\isacharunderscore}comp{\isacharunderscore}aux\ Ma\ sa\ {\isacharparenleft}tauclose{\isacharunderscore}step\ Ma\ sa\ xa{\isacharparenright}\ {\isacharequal}\isanewline
\ \ \ \ \ \ \ \ \ \ \ \ \ lfp\ {\isacharparenleft}tauclose{\isacharunderscore}step\ Ma\ sa{\isacharparenright}{\isachardoublequoteclose}\isanewline
\ \ \ \ \ \isakeyword{and}\ \ \ \ {\isachardoublequoteopen}xa\ {\isasymsubseteq}\ lfp\ {\isacharparenleft}tauclose{\isacharunderscore}step\ Ma\ sa{\isacharparenright}{\isachardoublequoteclose}\isanewline
\ \ \isacommand{from}\isamarkupfalse%
\ this\ \isacommand{show}\isamarkupfalse%
\ {\isachardoublequoteopen}tauclose{\isacharunderscore}comp{\isacharunderscore}aux\ Ma\ sa\ xa\ {\isacharequal}\ lfp\ {\isacharparenleft}tauclose{\isacharunderscore}step\ Ma\ sa{\isacharparenright}{\isachardoublequoteclose}\isanewline
\ \ \isacommand{proof}\isamarkupfalse%
\ {\isacharparenleft}simp{\isacharcomma}rule{\isacharunderscore}tac\ ih{\isacharcomma}simp{\isacharparenright}\isanewline
\ \ \ \ \isacommand{assume}\isamarkupfalse%
\ {\isachardoublequoteopen}xa\ {\isasymsubseteq}\ lfp\ {\isacharparenleft}tauclose{\isacharunderscore}step\ Ma\ sa{\isacharparenright}{\isachardoublequoteclose}\isanewline
\ \ \ \ \isacommand{from}\isamarkupfalse%
\ this\ \isacommand{show}\isamarkupfalse%
\ {\isachardoublequoteopen}tauclose{\isacharunderscore}step\ Ma\ sa\ xa\ {\isasymsubseteq}\ lfp\ {\isacharparenleft}tauclose{\isacharunderscore}step\ Ma\ sa{\isacharparenright}{\isachardoublequoteclose}\isanewline
\ \ \ \ \isacommand{by}\isamarkupfalse%
\ {\isacharparenleft}subst\ lfp{\isacharunderscore}unfold{\isacharcomma}\ unfold\ tauclose{\isacharunderscore}step{\isacharunderscore}def{\isacharcomma}auto\ simp\ add{\isacharcolon}mono{\isacharunderscore}tauclose{\isacharunderscore}step{\isacharparenright}\isanewline
\ \ \isacommand{qed}\isamarkupfalse%
\isanewline
\isacommand{qed}\isamarkupfalse%
\endisatagproof
{\isafoldproof}%
\isadelimproof
\isanewline
\endisadelimproof
\isanewline
\isacommand{lemma}\isamarkupfalse%
\ tauclose{\isacharunderscore}comp{\isacharunderscore}sound\ {\isacharcolon}\isanewline
\ \ \isakeyword{shows}\ {\isachardoublequoteopen}tauclose{\isacharunderscore}comp\ M\ s\ {\isacharequal}\ tauclose\ M\ s{\isachardoublequoteclose}\isanewline
\isadelimproof
\endisadelimproof
\isatagproof
\isacommand{by}\isamarkupfalse%
\ {\isacharparenleft}unfold\ tauclose{\isacharunderscore}comp{\isacharunderscore}def{\isacharcomma}\ auto\ simp\ add{\isacharcolon}\ tauclose{\isacharunderscore}comp{\isacharunderscore}aux{\isacharunderscore}sound{\isacharparenright}%
\endisatagproof
{\isafoldproof}%
\isadelimproof
\endisadelimproof
\begin{isamarkuptext}%
\begin{theorem}\label{th:tauclose_comp_sound}
\mbox{}\newline
\isa{tauclose{\isacharunderscore}comp\ M\ s\ {\isacharequal}\ tauclose\ M\ s}
\end{theorem}

The proof is by functional induction on \isa{tauclose{\isacharunderscore}comp{\isacharunderscore}aux}.%
\end{isamarkuptext}%
\isamarkuptrue%
\isadelimtheory
\endisadelimtheory
\isatagtheory
\endisatagtheory
{\isafoldtheory}%
\isadelimtheory
\endisadelimtheory
\end{isabellebody}%

\section{Conclusions}\label{sec:conclusions}

This paper has presented a new kind of small-step semantics for imperative
programming languages, based on the zipper data structure. Our primary aim is
to show that this semantics has decisive advantages for abstracting
programming language semantics to automata. Even if the generated automata
have a great number of silent transitions, these can be removed.

The playground of our formalizations is proof assistants, in which SOS has
become a well-established technique for presenting semantics of programming
languages. In principle, our technique could be adapted to other formalization
tools like rewriting-based ones \cite{SerbanutaRM09}.

We are currently in the process of adopting this semantics in a larger
formalization from Java to Timed Automata
\cite{baklanova12:_abstr_verif_proper_real_time_java}. As most constructs
(zipper data structure, mapping to automata) are generic, we think that this
kind of semantics could prove useful for similar formalizations with other
source languages. The proofs (here carried out with the Isabelle proof
assistant) have a pleasingly high degree of automation that are in sharp contrast
with the index calculations that are usually required when naming automata
states with numbers. 

Renaming nodes from source tree locations to numbers is nevertheless easy to
carry out, see the code snippet provided on the web page
\cite{zipper_formalization} of this paper. For these reasons, we think that
the underlying ideas could also be useful in the context of compiler
verification, when converting a structured source program to a flow graph with
basic blocs, but before committing to numeric values of jump targets.


\bibliographystyle{plain}
\bibliography{main}

\end{document}